\newif\ifCHANGES
\let\ifCHANGES\iftrue

\documentclass[
    3p,
    sort&compress,
    times,
]{elsarticle}

\usepackage[utf8]{inputenc}

\usepackage{xparse}

\usepackage{microtype}

\usepackage{amsmath}
\usepackage{amssymb}
\usepackage{bm}
\usepackage{mathtools}
\usepackage{physics}
\usepackage{siunitx}

\usepackage{graphicx}
\usepackage{xcolor}

\usepackage{todonotes}

\usepackage{acronym}
\usepackage{hyperref}

\definecolor{edits}{RGB}{220,0,0}
\definecolor{strike}{RGB}{150,50,50}
\ifCHANGES
    \usepackage[normalem]{ulem}
    
    \NewDocumentCommand\STRIKE{+m}{{\color{strike}\sout{#1}}}
    
\else
    
    \NewDocumentCommand\STRIKE{+m}{}
    
\fi

\NewDocumentCommand\eg{}{e.\,g.}
\NewDocumentCommand\ie{}{i.\,e.}
\NewDocumentCommand\cf{}{cf.}

\NewDocumentCommand\eqperiod{}{\,\text{.}}
\NewDocumentCommand\eqcomma{}{\,\text{,}}
\RenewDocumentCommand\qcomma{}{\eqcomma\quad}

\RenewDocumentCommand\vec{m}{\bm{#1}}
\NewDocumentCommand\mat{m}{\bm{#1}}
\NewDocumentCommand\snabla{m}{\vnabla_{\!#1}}
\DeclarePairedDelimiterX\set[1]{\{}{\}}
    {#1}

\NewDocumentCommand\Nr{}{\ensuremath{N_\uparrow}}

\NewDocumentCommand\kEnse{}{\ensuremath{k_\mathrm{e}}}

\NewDocumentCommand\kMani{}{\ensuremath{k_\mathrm{m}}}

\NewDocumentCommand\kFloq{}{\ensuremath{k_\mathrm{F}}}

\NewDocumentCommand\Ws{}{\ensuremath{\mathcal{W}_\mathrm{s}}}
\NewDocumentCommand\Wu{}{\ensuremath{\mathcal{W}_\mathrm{u}}}

\NewDocumentCommand\mx{}{m_x}

\NewDocumentCommand\mz{}{m_z}
\NewDocumentCommand\Hk{}{\ensuremath{H_K}}
\NewDocumentCommand\Ms{}{\ensuremath{M_\mathrm{S}}}
\NewDocumentCommand\Han{}{\ensuremath{\vec{H}^\mathrm{an}}}
\NewDocumentCommand\Hd{}{\ensuremath{\vec{H}^\mathrm{d}}}
\NewDocumentCommand\Hext{}{\ensuremath{\vec{H}^\mathrm{ext}}}
\NewDocumentCommand\Hextz{}{\ensuremath{H^\mathrm{ext}_z}}
\NewDocumentCommand\Hphi{}{\ensuremath{H_\varphi}}
\NewDocumentCommand\Htheta{}{\ensuremath{H_\theta}}

\NewDocumentCommand\ignore{+m}{}

\journal{Communications in Nonlinear Science and Numerical Simulation}

\begin{document}

\begin{frontmatter}
    \title{Transition state dynamics of a driven magnetic free layer}

    \author[us]{Johannes Mögerle}
    \author[us]{Robin Schuldt}
    \author[us]{Johannes Reiff}
    \author[us]{Jörg Main}
    \address[us]{
        Institut für Theoretische Physik I,
        Universität Stuttgart,
        70550 Stuttgart, Germany
    }

    \author[jhuc,jhucbe]{Rigoberto Hernandez\texorpdfstring{\corref{cor}}{}}
    \ead{r.hernandez@jhu.edu}
    \cortext[cor]{Corresponding author}
    \address[jhuc]{
        Department of Chemistry,
        Johns Hopkins University,
        Baltimore, Maryland 21218, USA
    }
    \address[jhucbe]{
        Departments of Chemical \& Biomolecular Engineering,
        and Materials Science and Engineering,
        Johns Hopkins University,
        Baltimore, Maryland 21218, USA
    }

    \date{\today}

    \begin{abstract}
        Magnetization switching in ferromagnetic
        structures is an important process for technical applications such
        as data storage in spintronics, and therefore the determination
        of the corresponding switching rates becomes essential.
        We investigate a free-layer system
        in an oscillating external magnetic field
        resulting in an additional torque on the spin.
        The magnetization dynamics including inertial
        damping can be described by the phenomenological Gilbert equation.
        The magnetization switching between the two stable
        orientations on the sphere then requires the crossing of a
        potential region characterized by a moving rank-1 saddle.
        We adopt and apply recent extensions of transition state theory for
        driven systems to compute both the time-dependent and average
        switching rates of the activated spin system in the saddle region.
    \end{abstract}

\end{frontmatter}


\acrodef{BCM}{binary contraction method}
\acrodef{DoF}{degree of freedom}
\acrodefplural{DoF}{degrees of freedom}
\acrodef{DS}{dividing surface}
\acrodef{LD}{Lagrangian descriptor}
\acrodef{LLG}{Landau--Lifshitz--Gilbert}
\acrodef{LMA}{local manifold analysis}
\acrodef{NHIM}{normally hyperbolic invariant manifold}
\acrodef{TD}{time descriptor}
\acrodef{TST}{transition state theory}
\acrodef{TS}{transition state}


\section{Introduction}
\label{sec:introduction}

In recent years,
the promise of spintronics to emerging technological applications
has attracted growing interest
leading to extensive research efforts
in experimental~\cite{Schneider2004, Jiang2005, Cowburn2007,
    Shinjo2014, Maekawa2017, Monakhov2017, Khodadadi2020a, Liu2020a}
and theoretical physics~\cite{Wolf2001, Schneider2004, Jiang2005, Rocha2005,
    Adam2006, Chappert2007, Taniguchi2013b}.
The relative simplicity and accuracy of the
single-domain models
for ferromagnetic structures
has proven to be a popular choice for
characterizing such spintronics applications.
Specifically, these models describe
the macro spin-dynamics underlying
the Gilbert equation~\cite{Gilbert2004, Apalkov2005, Abert2019a}.
The landscape of the corresponding potential includes
two minima at the stable \emph{spin up} and \emph{spin down} positions
which are separated by a rank-1 saddle
in certain configurations~\cite{LiZhang2003, LiZhang2004}.
The typical goal in spintronics applications is to achieve and control
the magnetization switching within a target timescale---viz., a
specified rate.
This can be achieved, for example,
through application of a spin torque~\cite{Slonczewski1996a}.
An alternative approach is microwave-assisted magnetic recording%
---more specifically, microwave-assisted switching~\cite{
    Zhu2008a, Okamoto2012a, Taniguchi2014b, Suto2015a}---%
where a microwave field perpendicular to the easy axis is used
in conjunction with a static external field along the easy axis
in order to facilitate the magnetization switching.
Multiple variations of this scheme have been proposed~\cite{
    Barros2011a, Barros2013a, Klughertz2015a, Taniguchi2016a},
some of which rely solely on rotating AC fields
perpendicular to the easy axis~\cite{Rivkin2006a, Taniguchi2015a}.
In this paper, we focus on a single AC field along the easy axis
without any static external fields.

In chemical reactions, the transition from reactants to products is
typically marked by a barrier region with a rank-1 saddle that has
exactly one unstable direction called the reaction coordinate, while
the remaining orthogonal modes are locally stable and are associated
with other bound internal motions.
The dynamical crossing of a rank-1 saddle in such chemical systems can
be described by \ac{TST}~\cite{eyring35, wigner37, pitzer, pech81,
truh96, peters14a, wiggins16}, which then allows for the calculation
of rate constants and the flux.
However, \ac{TST} is not restricted to chemical reactions as it has
been applied in many other fields, including, \eg, atomic
physics~\cite{Jaffe00}, solid state physics~\cite{Jacucci1984},
cluster formation~\cite{KomatsuzakiBerry99a, KomatsuzakiBerry02},
diffusion dynamics~\cite{toller, voter02b}, and
cosmology~\cite{Oliveira02, Jaffe02, hern20i}.
Notably, the theory has also been extended
to time-dependent driven systems~\cite{hern08d}.
Although originally framed
using perturbation theory~\cite{hern93b, Uzer02, Waalkens04b, Waalkens13},
the requisite locally recrossing-free \ac{DS}
and instantaneous decay rates
in \ac{TST} can now be obtained
with more generally-applicable
methods~\cite{hern14f, hern17h, hern19a, hern19e, hern20m}
as employed here.

Thus, the central result of this paper is the demonstration of the
applicability of
\emph{time-dependent} \ac{TST} to characterize
the dynamical crossing of a macrospin across a
time-dependent rank-1 saddle
using the recent advances cited above.
In the language of \ac{TST}, the \emph{spin up} and \emph{spin down}
regions can be interpreted as \emph{reactants} and \emph{products},
and the magnetization switching corresponds to the ``chemical'' reaction.
An important difference between the previous
systems to which \ac{TST} has so far been applied,
and the ferromagnetic systems
described by the Gilbert equation
lies in the geometry of the phase-space structure.
Typically, a Hamiltonian system with $d$ degrees of freedom is described by a
$(2 d)$-dimensional phase space with $d$ coordinates and $d$ associated
velocities or momenta.
The Gilbert equation, however, is a first-order differential equation
for the dynamics of the magnetic moment on a sphere, \ie, there are no
independent velocities or momenta.
Therefore, the dynamics is effectively that of a one \ac{DoF} system~%
\cite{Landau1935, Gilbert1956, Gilbert2004, Ciornei2010}.
Nevertheless,
within this domain a \ac{DS} can be associated
with the neighborhood of the rank-1 saddle.
In analogy to chemical reactions,
we conjecture that
the reactive flux across this \ac{DS}
is associated with the decay rate of the spin flip.
In this context, the reactive flux is that of all the trajectories
that are reactants (viz., \emph{spin up}) in the infinite past
and products (viz., \emph{spin down}) in the infinite future.
In transition state theory, the reactive flux is approximated
by the sum of the positive velocities (headed in the direction of the product)
over the surface, and it is exact if no trajectory recrosses the \ac{DS}.

We show that recent extensions of \ac{TST} for systems with
time-dependent moving saddles~\cite{hern14f,hern17h,hern19a,hern19e}
can indeed be applied to
a ferromagnetic single-domain system with a
two-dimensional phase space describing the orientation of the magnetic
moment on the sphere and the dynamics following the Gilbert equation.
The system can even be driven by a time-dependent external magnetic field.
The free-layer system and the applied methods
are introduced in Sec.~\ref{sec:theory}.
The applicability of \ac{TST} relies on the fact that for any time $t$,
the two-dimensional phase space exhibits a stable and unstable
manifold, which intersect in a point on the \ac{NHIM}.
A locally recrossing-free \ac{DS} separating the \emph{spin down} and
\emph{spin up} regions in phase space can be attached to this point.
The time-dependent moving points of the \ac{NHIM} form the \ac{TS} trajectory,
which is a periodic orbit when the free-layer system is
driven by an oscillating magnetic field.

The \ac{TS} trajectories are the starting point for the calculation of
rate constants, and the characterization of the magnetization switching.
Through application of the ensemble method and the \ac{LMA} developed in
Ref.~\cite{hern19e}, we obtain the time-dependent instantaneous rates
along \ac{TS} trajectories at various amplitudes and frequencies of
the driving external magnetic field.
We also find in Sec.~\ref{sec:results}
that the time-averaged rates along the \ac{TS} trajectories
depend significantly on the external driving.


\section{Theory and methods}
\label{sec:theory}

\begin{figure}
    \includegraphics[width=\columnwidth]{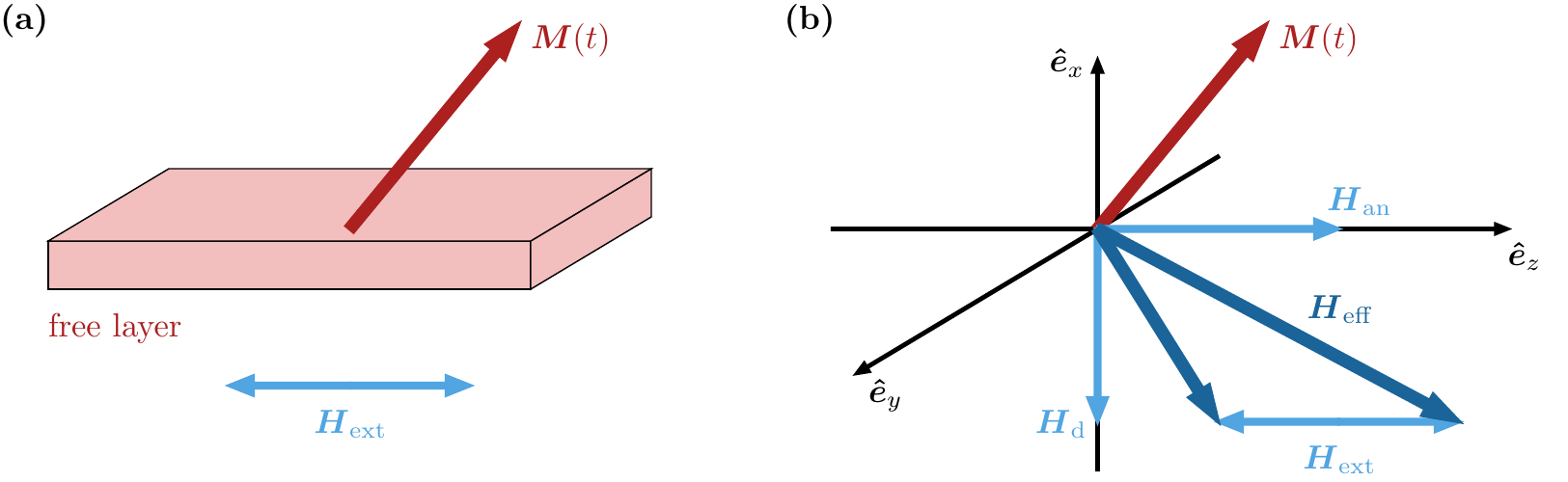}
    \caption{%
        (a)~Schematic of a magnetic single-domain layer
        with variable magnetization $\vec{M}$ (free layer)
        in an external magnetic field \Hext.
        (b)~Magnetic field components governing
        the evolution of the free layer's magnetization $\vec{M}$.
        The effective field $\vec{H}$ acting upon $\vec{M}$ consists of
        the demagnetization field $\Hd = -\Ms \mx \vu*{e}_x$,
        the magnetocrystalline anisotropy field $\Han = \Hk \mz \vu*{e}_z$,
        and the external driving $\Hext = \Hextz \sin(\omega t) \vu*{e}_z$.
        The material's easy axis is aligned with the $z$-axis.}
    \label{fig:config}
\end{figure}

Here, we briefly discuss our model (\cf\ Fig.~\ref{fig:config})
motivated by a free-layer system~\cite{LiZhang2003}
and present the equations, which describe
the spin dynamics of this model including external driving.
Then we introduce the basic ideas of \ac{TST} and the methods, which
will be applied for the computation of the instantaneous and average
rates of the magnetization switching.


\subsection{Spin dynamics in a driven free-layer model}
\label{sec:model}

The model addressed here
is based on a magnetic single-domain layer
with variable magnetization $\vec{M}$, known as a \emph{free layer}.
This layer is modeled in analogy to
Stoner and Wohlfarth~\cite{Stoner1948a, Tannous2008a}
including a demagnetization field for a thin film
(shape anisotropy)~\cite{Zhang2004a, Tannous2008a}.
A periodic external magnetic field is added to drive the magnetization.
This field is intended as
a generic placeholder for some externally applied torque%
---\eg, certain types of spin torque~\cite{Manchon2008a}
such as the one stated below---%
and must not necessarily be realized by a magnetic coil or antenna.

For a classical description of the spin system we start from the
Gilbert equation~\cite{Gilbert1956, Gilbert2004}
\begin{equation}
    \label{eq:dotm_implicit}
    \dot{\vec{M}} = -\gamma \vec{M} \cp \vec{H}
        + \frac{\alpha}{\Ms} \vec{M} \cp \dot{\vec{M}}
\end{equation}
to describe the motion of a magnetic moment $\vec{M} = -\gamma \vec{S}$,
with $\vec{S}$ the spin, $\gamma$ the gyromagnetic ratio, and
$\Ms = \abs{\vec{M}}$ the saturation magnetization
[\cf\ Fig.~\ref{fig:config}(b)].
The magnetic moment $\vec{M}$ is damped by a strength proportional to
the coefficient $\alpha$ and can be driven by the effective magnetic
field $\vec{H}$.
Because the velocity $\dot{\vec{M}}$ is orthogonal to $\vec{M}$, the
length of the magnetic moment is conserved and therefore we can write
$\vec{M} = \Ms \vec{m}$ with $\abs{\vec{m}} = 1$ and $\vec{m}$ being
dimensionless.

The implicit differential equation~\eqref{eq:dotm_implicit} can be
brought to an explicit form.
Substituting $\dot{\vec{M}}$ on the right-hand side of
Eq.~\eqref{eq:dotm_implicit} with the equation itself and using the
relation $\vec{M} \cp (\vec{M} \cp \dot{\vec{M}})
= (\vec{M} \vdot \dot{\vec{M}}) \vec{M} - \vec{M}^2 \dot{\vec{M}}
= - \Ms^2 \dot{\vec{M}}$ as well as $\vec{M} = \Ms \vec{m}$
we obtain the \ac{LLG} equation~\cite{Gilbert2004,Abert2019a}
\begin{equation}
    \label{eq:dotm}
    \dot{\vec{m}} = -\frac{\gamma}{1 + \alpha^2} \vec{m}
        \cp \qty[\vec{H} + \alpha \qty(\vec{m} \cp \vec{H})]
    \eqperiod
\end{equation}

Here, we investigate the motion of a magnetic moment
in a free-layer model
described by the potential~\cite{Zhang2004a}
\begin{equation}
    \label{eq:U}
    U = \frac{\Ms^2}{2} \mx^2
        - \frac{\Ms\Hk}{2} \mz^2
        - \Ms \Hextz \sin(\omega t) \mz
    \eqcomma
\end{equation}
where \Hk\ is the anisotropy constant of the free layer.
A magnetization switching induced by an additional torque
modifying the dynamics of Eq.~\eqref{eq:dotm}, can in principle be
achieved by various ways~\cite{
    Zhu2008a, Okamoto2012a, Taniguchi2014b, Suto2015a,
    Barros2011a, Barros2013a, Klughertz2015a, Taniguchi2016a,
    Rivkin2006a, Taniguchi2015a}.
For the description of spin torque in a pinned-layer system,
Slonczewski introduced an additional term to the standard Gilbert equation,
depending on the polarization of the pinned layer~\cite{Slonczewski1996a}.
In this model, the spin torque is proportional to the applied electron
current $I$ flowing trough the pinned layer and, thus, can in
principle become oscillating if an AC-source is
used~\cite{McMichael1999a, Zhu2020a}.
While this specific type of spin torque
cannot be represented purely by an additional magnetic field term,
others---\eg, Manchon and Zhang~\cite{Manchon2008a}---%
have suggested spin torques that can.
Due to the fact that the influence of
some spin torques
can be reformulated as an additional effective field acting on the
spin dynamics~\cite{Apalkov2005b, Manchon2008a},
we directly add our
applied field expression into the effective field, leading to significant
simplifications
~\cite{Abert2019a}.
The last term in Eq.~\eqref{eq:U} describes such an oscillating
external magnetic field in $z$ direction with amplitude \Hextz\ and
frequency $\omega$.
The effective magnetic field then reads
\begin{equation}
    \label{eq:H-ausfuehrlich}
    \vec{H} = -\frac{1}{\Ms} \snabla{\vec{m}} U
        = \mqty(- \Ms \mx \\ 0 \\ \Hk \mz + \Hextz \sin(\omega t))
    \eqperiod
\end{equation}

For the free-layer system with
parameters 
based on Refs.~\cite{Taniguchi2014b, Taniguchi2015a, Taniguchi2016a}
the saturation magnetization and the gyromagnetic ratio read
\begin{equation}
    \Ms = \SI{1e6}{\A\per\m}
    \qand
    \gamma = \SI{2.217e5}{\m\per\A\per\s}
    \eqcomma
\end{equation}
respectively.
Using these values as units, we can set $\Ms = \num{1}$ and
$\gamma = \num{1}$ for computations with dimensionless parameters.
In the following, we choose
\begin{equation}
    \label{eq:static_parameters}
    \Ms = \num{1} \qc
    \gamma = \num{1} \qc
    \alpha = \num{0.01} \qc
    \Hk = \num{0.5} \eqcomma
\end{equation}
and an external magnetic field with amplitude and frequency
\begin{equation}
    \label{eq:driving_parameters}
    \Hextz = \num{0.15}
    \qand
    \omega = \pi / 8
\end{equation}
as reference parameters, if not stated otherwise.
This corresponds to
$\Hk = \SI{5e5}{\A\per\m}$,
$\Hextz = \SI{1.5e5}{\A\per\m}$, and
$\omega / 2 \pi = \SI{13.86}{\GHz}$
in the problem defined in Ref.~\cite{Najafi2009a} with the standard
material parameters of permalloy.
This applied field and frequency are well in the range
of typical experimental conditions.

\begin{figure}
    \includegraphics[width=0.31\columnwidth]{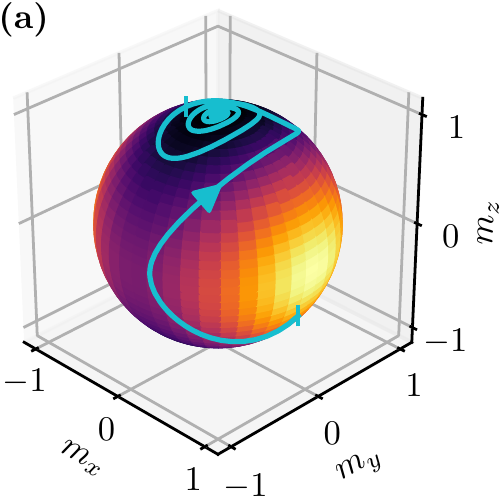}
    \hfill
    \includegraphics[width=0.67\columnwidth]{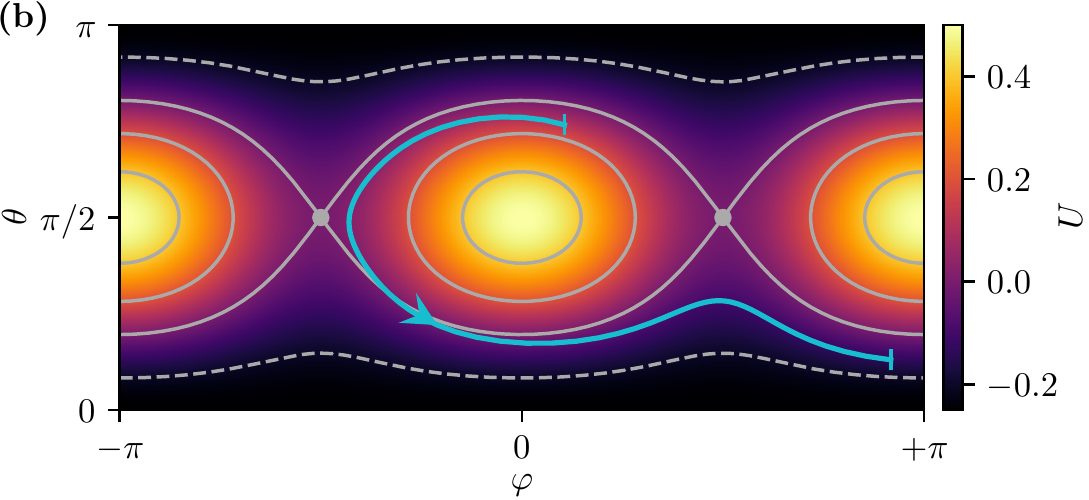}
    \caption{%
        The free-layer potential~\eqref{eq:U}
        (a)~on the sphere and
        (b)~in the $(\varphi, \theta)$ plane.
        The saddle points at $\theta = \pi / 2$ and $\varphi = \pm \pi / 2$
        mark the regions of the \acs{TS}, which must be crossed for
        the magnetization switching.
        A typical trajectory with higher friction $\alpha = \num{0.1}$,
        propagated without external driving
        from a spin-down state to a spin-up state, is
        shown in cyan (or light gray in print) in both panels.
        Vertical markers highlight the part of the trajectory shown in (b).}
    \label{fig:potential}
\end{figure}

To take advantage of the symmetry of the system one can transform the
\ac{LLG} equation~\eqref{eq:dotm} in spherical angular
coordinates $\theta$ and $\varphi$, \ie,
\begin{equation}
    \label{eq:dotm-sp}
    \dot{\theta} = \frac{\gamma}{1 + \alpha^2}
        \qty(\Hphi + \alpha \Htheta)
    \qcomma
    \dot{\varphi} = \frac{\gamma}{1 + \alpha^2} \frac{1}{\sin\theta}
        \qty(-\Htheta + \alpha \Hphi)
    \eqcomma
\end{equation}
for $\theta \notin \set{0, \pi}$ and with the projections of the
effective field to the basis vectors,
$\Htheta = \vec{H} \vdot \vu*{e}_\theta$ and
$\Hphi = \vec{H} \vdot \vu*{e}_\varphi$.
The potential~\eqref{eq:U} on the sphere for the spin system without
external driving is illustrated in Fig.~\ref{fig:potential}(a).
The reduced equations of motion~\eqref{eq:dotm-sp} for the two
spherical angular variables describe a two-dimensional problem with the
potential $U(\theta, \varphi)$ presented in Fig.~\ref{fig:potential}(b).
Two energy minima are located at the poles with $\theta = 0$ and
$\theta = \pi$, which make the potential~\eqref{eq:U}
a perfect candidate to observe magnetization switching.
At the equator $\theta = \pi / 2$ there are two local energy maxima
for $\varphi = 0$ and $\varphi = \pi$ and two rank-1 saddles for
$\varphi = -\pi / 2$ and $\varphi = +\pi / 2$.

A typical trajectory without external driving illustrating the
magnetization switching is drawn in both panels (a) and (b) of
Fig.~\ref{fig:potential}.
It starts on the \emph{spin down} side of the sphere ($\theta > \pi / 2$),
crosses the saddle region of the potential near $\theta = \pi / 2$,
$\varphi = -\pi / 2$ and approaches the \emph{spin up} position
($\theta \approx 0$) on a spiral caused by the damping term in the
Gilbert equation~\eqref{eq:dotm_implicit}.
We are interested in spin-flip processes crossing the regions close to
one of the rank-1 saddles, and investigate in the following, without
loss of generality, spin flips crossing the rank-1 saddle near
$\varphi = +\pi / 2$.


\subsection{Transition state theory}
\label{sec:tst}

The free-layer system, described by the potential~\eqref{eq:U},
features a rank-1 saddle point at $\theta = \varphi = \pi / 2$, as shown
in Fig.~\ref{fig:potential}(b).
This saddle can act as a bottleneck of the spin dynamics, which makes
it a candidate for the application of \ac{TST}
models~\cite{eyring35, wigner37, pech81, wiggins16}.
In typical scenarios for a chemical reaction,
a one-dimensional reaction path%
---\eg, the minimum energy path~\cite{Glasstone1941}---%
characterizes the progress of the reaction.
A rank-1 saddle point separates reactants from products
along this unstable mode,
and can be used to naively characterize the flux and associated reaction rate.
In this context, it acts as a \ac{TS}.
In higher dimensions, the other degrees of freedom are stable and
are referred to as orthogonal modes.
More generally, the \ac{TS} marks the transition
between reactants and products through the location of a
\ac{DS}.
Here, we apply \ac{TST} to a magnetization switching in the
free-layer system%
---\eg, from the ``reactant'' state \emph{spin up}
to the ``product'' state \emph{spin down}---caused by a time-dependent
driving of the system via an external magnetic field.
To achieve this aim, we resort to recent extensions of \ac{TST} to
time-dependent driven systems~\cite{hern17h, hern19a, hern19e, hern20m}.


\subsubsection{Phase-space structure and TS trajectory}
\label{sec:phase_space}

In the free-layer system introduced above,
the magnetization switching
is related to a change in the $\theta$
coordinate---\eg, from $\theta \gtrsim 0$ to $\theta \lesssim \pi$
in an \emph{up} to \emph{down} spin state.
In applying \ac{TST} to resolve the activated dynamics of a spin,
it thus appears natural to take the
angle $\theta$ as the reaction coordinate and $\varphi$ as an
orthogonal mode.
However, an important difference between the spin system described by
the equations of motion in~\eqref{eq:dotm-sp},
and systems typically addressed by \ac{TST}
requires some considerations, discussed below, to make the analogy complete.

In a chemical or mechanical system with $d$ degrees of freedom the
dynamics is typically described by $d$ second-order differential
equations for the coordinates or, in the Hamilton formalism, by $2 d$
first-order differential equations for the coordinates and canonical
momenta in the $2 d$-dimensional phase space.
In the spin system, the \ac{LLG} equation results in the
\emph{first-order} differential equations~\eqref{eq:dotm-sp} for the
two coordinates $\theta$ and $\varphi$, \ie, there are no canonical
momenta $p_\theta$ and $p_\varphi$, which belong to these coordinates.
Nevertheless, \ac{TST} can be applied to this system.
The crucial point is that the two-dimensional phase space of the spin
system consisting of the two coordinates $\theta$ and $\varphi$ is
treated in formal mathematical analogy to the two-dimensional phase
space of a one \ac{DoF} Hamiltonian system with a reaction coordinate
and the corresponding canonical momentum.

\begin{figure}
    \includegraphics[width=0.5\columnwidth]{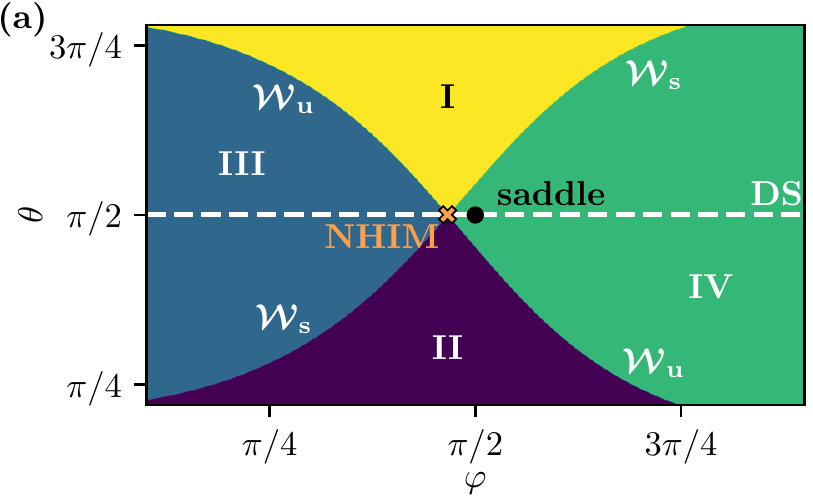}
    \hfill
    \includegraphics[width=0.5\columnwidth]{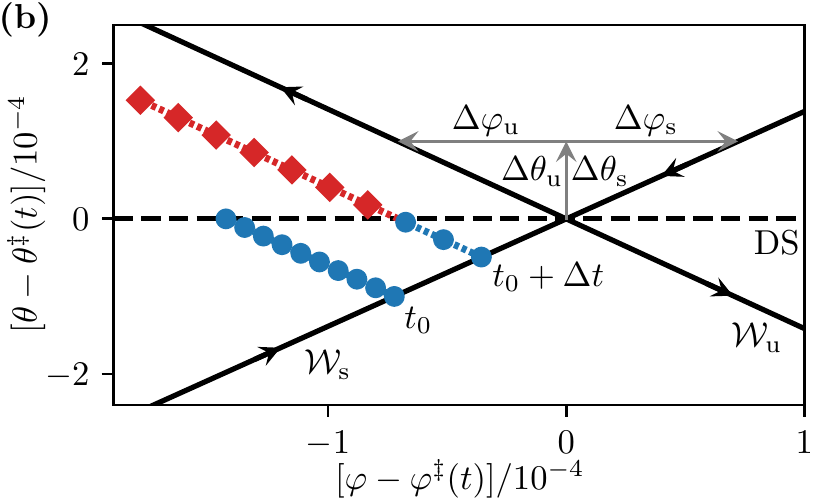}
    \caption{%
        (a)~Phase-space structure of the driven free-layer system
        introduced in Sec.~\ref{sec:model}.
        The stable and unstable manifolds, \Ws\ and \Wu\
        separate four different regions marked $(I)$--$(IV)$ (see text).
        The intersection of the manifold's closures forms the \acs{NHIM}.
        The \acs{DS} attached to this point separates the \emph{spin up}
        and \emph{spin down} regions in phase space.
        The external driving causes the \acs{NHIM}
        to detach from the saddle point.
        (b)~Schematic of the geometric structure that underlies the
        rate constant expressions summarized in Sec.~\ref{sec:rates}.
        Initially, an equidistant spin ensemble connecting \Ws\ on the
        reactant side with the \acs{DS} parallel to \Wu\ is generated.
        Upon time propagation, parts of the ensemble undergo
        spin flips as they move through the \acs{DS}.
        The resulting ensemble is still equidistant, parallel to \Wu,
        and connected to \Ws.}
    \label{fig:tst}
\end{figure}

The phase-space structure of the driven spin system in the
vicinity of the rank-1 saddle at a given time $t$ is
illustrated in Fig.~\ref{fig:tst}(a).
Note that the reaction coordinate $\theta$ is the ordinate and
$\varphi$ the abscissa, which differs from corresponding presentations
in Refs.~\cite{hern17h,hern18g,hern19a,hern19e,hern20m}, where the
reaction coordinate is chosen as the abscissa and the corresponding
velocity along the ordinate.
The stable and unstable manifolds \Ws\ and \Wu\ separate four
different regions, where
$(I)$ the spin stays \emph{down},
$(II)$ the spin stays \emph{up},
$(III)$ the spin switches from \emph{up} to \emph{down}, and
$(IV)$ the spin switches from \emph{down} to \emph{up},
when the system is propagated backwards and forwards in time.
One subtlety regarding the time propagation of the spins should be
noted:
Due to the damping of the magnetic field by the term proportional to $\alpha$
in Eq.~\eqref{eq:dotm_implicit} the spin without external driving
always moves towards a potential minimum, \ie, the \emph{spin up} or
\emph{spin down} position when propagated forwards in time.
However, it moves towards one of the potential maxima located at
$\theta = \pi / 2$, $\varphi = 0$ or $\theta = \pi / 2$, $\varphi = \pi$
(see Fig.~\ref{fig:potential}) when propagated backwards.
Therefore, appropriate cutoffs for the propagation of trajectories
must be introduced to obtain the correct classification to one of the
regions $(I)$--$(IV)$ in Fig.~\ref{fig:tst}(a).
Failing to do so can lead to visible artifacts,
or it can cause the classification algorithm to not terminate.
Similar problems in dissipative chemical systems have been discussed
in Ref.~\cite{hern16i}.
In our case, we have found $\num{0.1} \pi < \varphi < \num{0.9} \pi$
to yield reliable results.

The intersection of the stable and unstable manifold
is a point $(\varphi^\ddagger, \theta^\ddagger)$ on the \ac{NHIM}.
Such points do not leave the saddle region when propagated forwards or
backwards in time.
Therefore, these points describe spins that reside permanently
in an unstable intermediate state roughly in $x$ direction
that is neither \emph{spin up} nor \emph{spin down}.
Note that for driven systems the points of the \ac{NHIM} in general do
not coincide with the time-dependent position of the saddle marked by
the black point in Fig.~\ref{fig:tst}(a).
The line with constant angle $\theta = \theta^\ddagger$ represents a
recrossing-free \ac{DS}, which separates the ``reactants'' and
``products'' in \ac{TST}, \ie, a spin with
$\theta < \theta^\ddagger$ is \emph{spin up} and a spin with
$\theta > \theta^\ddagger$ is \emph{spin down}.
In case of periodic driving of the spin system by a time-dependent
external magnetic field, the points on the \ac{NHIM} follow a periodic
orbit with the same period as the external driving.
This orbit is called the \ac{TS} trajectory, and is of fundamental
importance for the computation of rate constants.

For the numerical construction of the \ac{NHIM}, we resort to the \ac{BCM}
introduced in Ref.~\cite{hern18g}.
For a given time $t$, the algorithm in the \ac{BCM} is initialized by
defining a quadrangle with each of its corners lying exclusively
within one of the four regions in the $(\varphi, \theta)$ plane shown
in Fig.~\ref{fig:tst}(a).
In each iterative step, we first determine an edge's midpoint.
Then, the adjacent corner corresponding to the same region as that midpoint
is moved to the midpoint's position.
By repeating this interleaved bisection procedure in turn for all
edges, the quadrangle successively contracts and converges towards the
intersection of the stable and unstable manifolds, \ie, a point on
the \ac{NHIM}.
This method is numerically very effective and efficient for systems
such as the one addressed here.


\subsubsection{Decay rates}
\label{sec:rates}

Three different methods for calculating decay rates in driven systems
have recently been introduced and applied in the
literature~\cite{hern19e, hern20m}.
Here, we adopt these methods with appropriate modifications
for the free-layer system.
The resulting decay rates
are a measure of the instability of specific trajectories near the saddle.
They differ significantly from the Kramers rate~\cite{pgh89}
used in the theory of chemical reactions
but nevertheless provide insight about the rate process.

\paragraph{Ensemble method}

The conceptually simplest method for calculating decay rates
\kEnse\ is by means of propagation of an ensemble.
In analogy to Ref.~\cite{hern19e} we identify a line segment
parallel to the unstable manifold that satisfies the property:
it lies on the reactant side between the stable manifold and the
\ac{DS} at a distance that is small enough to allow for linear
response and large enough to suppress numerical instability.
At $t = t_0$, a spin ensemble is placed on this line as illustrated by
blue dots in Fig.~\ref{fig:tst}(b) and propagated in time
to yield a time-dependent \emph{spin up} population $\Nr(t)$.
The ensemble at time $t = t_0 + \Delta t$ is marked in
Fig.~\ref{fig:tst}(b) by red and orange dots.
Spins, which have crossed the \ac{DS} (red dots) are \emph{spin down}
and thus cause a decrease of the population $\Nr(t)$ (see the orange
dots) with increasing time.
In principle, one can now obtain a reaction rate constant \kEnse\ by
fitting an exponential decay $\Nr(t) \propto \exp[-\kEnse (t - t_0)]$
to the \emph{spin up} population.
This, however, is not possible in all systems because the decay in
$\Nr(t)$ can be nonexponential.
Instead, we use the more general approach described in Ref.~\cite{hern19e},
which involves examining the instantaneous decays
\begin{equation}
    \label{eq:ratemethods/k_e}
    \kEnse(t) = -\frac{\dot N_\uparrow(t)}{\Nr(t)}
    \eqperiod
\end{equation}

\paragraph{Local manifold analysis}

The ensemble method is computationally expensive because it requires
the propagation of a large number of spins for sufficiently long time.
An alternative method, called the \ac{LMA},
can be used to obtain
instantaneous spin-flip rates purely from the geometry of the stable
and unstable manifolds in phase space.
The \ac{LMA} is based on the observation that the equations of
motion~\eqref{eq:dotm-sp} can be linearized in the local vicinity of a
trajectory $\vec{m}^\ddagger(t)$
on the \ac{NHIM} with the Jacobian
\begin{equation}
    \label{eq:J}
    \mat{J}(t) = \eval{\pdv{(\dot{\theta}, \dot{\varphi})}{(\theta, \varphi)}}
        _{\vec{m}^\ddagger(t)}
    \eqperiod
\end{equation}
With the (not necessarily normalized) directions of the stable and
unstable manifolds \Ws\ and \Wu\ at time $t$ given by
$(\Delta\varphi_\mathrm{s}, \Delta\theta_\mathrm{s})$ and
$(\Delta\varphi_\mathrm{u}, \Delta\theta_\mathrm{u})$ with
$\Delta\theta_\mathrm{s} = \Delta\theta_\mathrm{u}$,
as marked in Fig.~\ref{fig:tst}(b), and using the linearization of
the equations of motion~\eqref{eq:dotm-sp} with the Jacobian~\eqref{eq:J}
for the propagation of the spin ensemble, we finally obtain an
analytical expression for the instantaneous rates of the magnetization
switching
\begin{equation}
    \label{eq:ratemethods/k_m}
    \kMani(t)
    = -\frac{\dot{N}_\uparrow(t)}{\Nr(t)}
    = -\lim_{\Delta t \to 0}
        \frac{\Nr(t + \Delta t) - \Nr(t)}{\Delta t \Nr(t)}
    = \eval{\pdv{\dot{\theta}}{\varphi}}_{\vec{m}^\ddagger(t)}
        \qty(\frac{\Delta\varphi_\mathrm{u}}{\Delta\theta_\mathrm{u}}
            - \frac{\Delta\varphi_\mathrm{s}}{\Delta\theta_\mathrm{s}})
    \eqperiod
\end{equation}
These rates can be calculated independently at different times $t$,
which allows for computations in parallel.
Note that $\Delta\varphi_\mathrm{s} / \Delta\theta_\mathrm{s}$ and
$\Delta\varphi_\mathrm{u} / \Delta\theta_\mathrm{u}$ are the inverse
slopes of the stable and unstable manifolds \Ws\ and \Wu\ in
Fig.~\ref{fig:tst}(b), and thus the instantaneous rate $\kMani(t)$
in Eq.~\eqref{eq:ratemethods/k_m} is mainly determined by the
difference of these two inverse slopes.
This differs from, \eg, Ref.~\cite{hern19e}, where the instantaneous rate
is related to the slopes of the stable and unstable manifolds; the
inverse slopes in Eq.~\eqref{eq:ratemethods/k_m} occur because the
reaction coordinate $\theta$ is not the abscissa but the ordinate in
Figs.~\ref{fig:tst}(a) and \ref{fig:tst}(b).
As discussed above, the angles $\theta$ and $\varphi$ are not
canonical variables as is typical in applications of \ac{TST}
to systems with Hamiltonian dynamics~\cite{hern19e, hern20m, hern20i}.
This manifests in a nontrivial and time-dependent prefactor
$\eval*{(\pdv*{\dot{\theta}}{\varphi})}_{\vec{m}^\ddagger(t)}$
in Eq.~\eqref{eq:ratemethods/k_m}, which is an element of the
Jacobian~\eqref{eq:J}.
In the limiting case of a Cartesian reaction coordinate $x$ with canonical momentum
$p = m \dot{x}$ (where $m$ is the particle mass),
the corresponding element of
the Jacobian reduces to a constant $\pdv*{\dot{x}}{p} = 1 / m$~\cite{hern19e}.

\paragraph{Floquet method}

The average decay rates \kFloq\ across time-dependent barriers
can also be obtained directly using
a Floquet stability analysis~\cite{hern14f, hern19e}.
While this method is computationally much cheaper than the
ensemble method and the \ac{LMA},
it cannot yield instantaneous rates.

To obtain the time-independent rate constant \kFloq\ for a given
\ac{TS} trajectory on the \ac{NHIM}, we linearize the equations of
motion using the Jacobian~\eqref{eq:J}.
By integrating the differential equation
\begin{equation}
    \dot{\mat{\sigma}}(t) = \mat{J}(t) \mat{\sigma}(t)
    \qq{with} \mat{\sigma}(0) = \mat{1}
    \eqcomma
\end{equation}
we then obtain the system's fundamental matrix $\mat{\sigma}(t)$.
When considering trajectories with period $T$,
$\mat{M} = \mat{\sigma}(T)$ is called the monodromy matrix.
Its eigenvalues $m_\mathrm{u}$ and $m_\mathrm{s}$, termed Floquet multipliers,
can be used to determine the Floquet rate constant
\begin{equation}
    \label{eq:ratemethods/k_f}
    \kFloq = \frac{1}{T} \qty(\ln\abs{m_\mathrm{u}} - \ln\abs{m_\mathrm{s}})
    \eqperiod
\end{equation}
As shown below, the Floquet rate constant $\kFloq$ agrees perfectly
with the instantaneous rates $\kEnse(t)$ and $\kMani(t)$ when the latter two are
averaged over one period $T$ of the \ac{TS} trajectory.


\section{Results and discussion}
\label{sec:results}

\begin{figure}
    \includegraphics[width=\columnwidth]{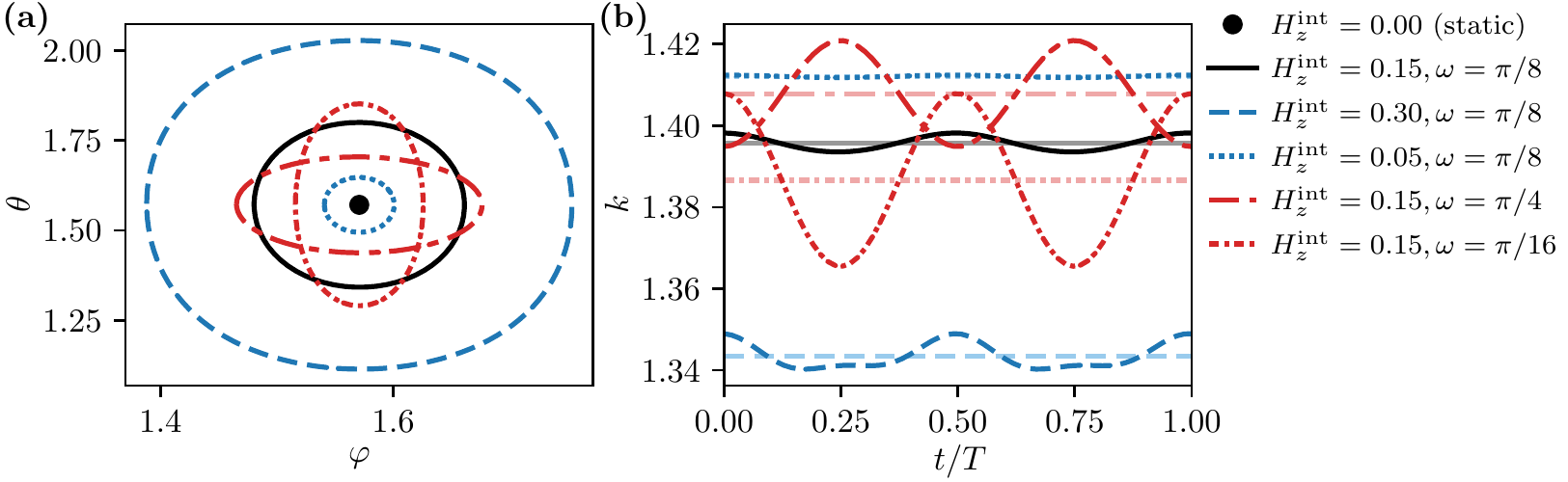}
    \caption{%
        (a)~A selection of \acs{TS} trajectories
        of the free-layer system
        with the potential~\eqref{eq:U}
        described by the \ac{LLG} equation~\eqref{eq:dotm}.
        The static \acs{TS} trajectory
        without external magnetic field ($\Hextz = 0$)
        is marked by a black dot at $\theta = \varphi = \pi / 2$.
        The \acs{TS} trajectory with the reference parameters
        given in Eqs.~\eqref{eq:static_parameters}
        and~\eqref{eq:driving_parameters}
        is shown as solid black line.
        \acs{TS} trajectories with driving parameters
        deviating from Eq.~\eqref{eq:driving_parameters}
        are drawn with colored dash or dash-dotted lines.
        The elliptical shape and orientation of the \acs{TS}
        trajectories depends strongly on the driving
        by the oscillating external magnetic field.
        (b)~Instantaneous rates (dark lines) and mean rates
        (pale lines)
        for some of the periodic \acs{TS} trajectories shown in (a).}
    \label{fig:dynamics}
\end{figure}

We now present and discuss the \ac{TS} trajectories and the related
instantaneous and averaged decay rates obtained for the
free-layer system with and without driving by an oscillating
external magnetic field.
\ac{TS} trajectories in the $(\varphi, \theta)$ phase space at various
amplitudes \Hextz\ and frequencies $\omega$
of the driving field
relative to the reference system of Eq.~\eqref{eq:driving_parameters}
are shown in Fig.~\ref{fig:dynamics}(a).
The static \ac{TS} trajectory without external driving is marked by a
black dot at $\theta = \varphi = \pi / 2$, which coincides with
the position of the static saddle in Fig.~\ref{fig:potential}.
When driven by an oscillating external field,
the \ac{TS} trajectories become periodic orbits
with the same period as the driving.
The elliptical shape and orientation of the orbits strongly depend on
the amplitude and frequency of the driving.
The black solid line marks the \acs{TS} trajectory with the
reference parameters given in Eqs.~\eqref{eq:static_parameters}
and~\eqref{eq:driving_parameters}.
The dashed lines mark \acs{TS} trajectories, where either the
amplitude \Hextz\ (blue lines) or the frequency $\omega$ (red lines)
deviates from these reference parameters.
For the chosen sets of parameters investigated here,
the driving frequency mostly affects the shape of the orbits,
whereas the driving amplitude has a large influence on the orbit size
while preserving the shape approximately.

Rate constants, which are related to the \ac{TS} trajectories
in Fig.~\ref{fig:dynamics}(a), have been computed with the methods
introduced in Sec.~\ref{sec:rates} and are shown in Fig.~\ref{fig:dynamics}(b).
The (dark lines mark the instantaneous rates
obtained by the \ac{LMA}
as functions of $t / T$ where $T = 2 \pi / \omega$ is the period of the
corresponding \ac{TS} trajectory.
As can be seen, the oscillation amplitude of the instantaneous rates
at high amplitude $\Hextz = 0.3$ of the driving field
(dark blue line)
is slightly higher than
that of the system with $\Hextz = 0.15$ ( black line).
This trend continues for $\Hextz = 0.05$,
where the oscillation is almost unnoticeable.
The pale lines present the averaged rate constants.
Here, the increase of the \Hextz\
from \num{0.15} (light gray line)
to \num{0.3} (pale blue line)
causes a significant decrease in the averaged rate constant.
The dark and pale red lines
in Fig.~\ref{fig:dynamics}(b)
mark the instantaneous and averaged decay rate
of the system at lower frequency $\omega = \pi / 16$ and
higher frequency $\omega = \pi  / 4$
of the oscillating magnetic field.
The instantaneous rate fluctuates
much stronger around the mean value.
The alternation in the strength of these fluctuations
is strong evidence for
a sign change in the modulation amplitude
around the reference frequency.
As mentioned above, the rate constants obtained as time averages of the
instantaneous rates over one period of the \ac{TS} trajectory agree
perfectly with the rate constants computed using the Floquet method.

\begin{figure}
    \centering
    \includegraphics[width=0.6\columnwidth]{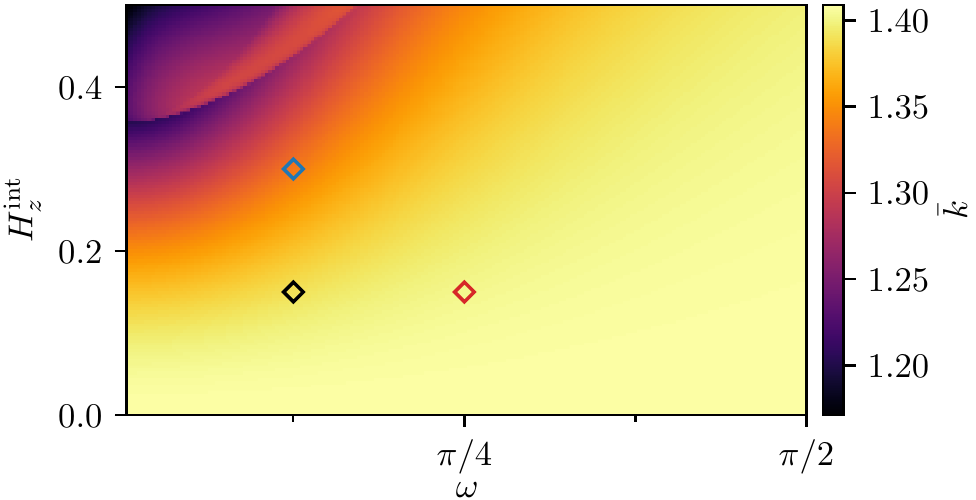}
    \caption{%
        Mean rates $\bar{k}$ as function of the frequency $\omega$
        and amplitude \Hextz\ of the external magnetic field.
        The diamonds mark the parameters used in Fig.~\ref{fig:dynamics}(b).}
    \label{fig:rates}
\end{figure}

Finally,
the dependence of the averaged
decay rate $\bar{k}$ on the amplitude \Hextz\ and
frequency $\omega$ of the magnetic field is reported
in Fig.~\ref{fig:rates}.
The diamonds mark the parameters of the \ac{TS} trajectories shown in
Fig.~\ref{fig:dynamics}(b).
A minimum in the rates lies near the corner with low frequencies
$\omega$ and high amplitudes \Hextz.


\section{Conclusion and outlook}
\label{sec:conclusion}

We have investigated magnetization switching in a ferromagnetic
free-layer system.
The dynamics of the magnetic moment is
described by the Gilbert equation~\eqref{eq:dotm_implicit}.
We have shown that \ac{TST} can be applied to its
two-dimensional phase space even though the Gilbert equation
does not have the expected
structure of a Hamiltonian system with coordinates and canonical
momenta.
We obtained the periodic \ac{TS} trajectories of the free-layer
system driven by an additional oscillating external magnetic field.
In turn, these form the basis
for the calculation of the instantaneous and averaged decay rates.
The rates significantly depend on the time-dependent driving, \ie,
the amplitude and frequency of the external magnetic field.
The magnetization switching can thus be controlled by the external
driving.

In this paper, we have assumed that the time derivative $\dot{\vec{m}}$
of the magnetic moment follows the magnetic field without relaxation,
as described by the Gilbert equation~\eqref{eq:dotm_implicit}.
In future work, the model for the free-layer system
could be extended by
taking into account relaxation of the spins~\cite{Fahnle2011}, which
requires one to enlarge the phase space from two to four dimensions.
\ac{TST} will then allow us to study the influence of the relaxation
on the decay rates.

Perhaps surprisingly, an increase in the field \Hextz\
mostly leads to a decrease in the mean rate
in Fig.~\ref{fig:rates}.
This is
perhaps a consequence of the intermediate friction
regime wherein the population of activated
spins---\ie, those that would go over the barrier---are
dampened by the dissipation.
Moreover, as the driving frequency increases,
the moving trajectory explores a wider oscillation potentially
averaging---and suppressing---the difference in the
curvatures associated with the stable and unstable directions
that contributes to the rate.
Resolution of this phenomenon remains a challenge
for future work.

In summary, this work suggests
that the application of
recent advances in locally nonrecrossing \ac{TST}
to magnetization
switching could be helpful in future work
addressing dynamics in spintronics.


\section*{Declaration of competing interest}

The authors declare that they have
no known competing financial interests or personal relationships
that could have appeared to influence the work reported in this paper.


\section*{CRediT authorship contribution statement}

\textbf{Johannes Mögerle:}
    Methodology,
    Software,
    Formal analysis,
    Investigation,
    Writing -- Original Draft.
\textbf{Robin Schuldt:}
    Methodology,
    Formal analysis,
    Investigation,
    Writing -- Original Draft.
\textbf{Johannes Reiff:}
    Methodology,
    Software,
    Validation,
    Resources,
    Data Curation,
    Writing -- Review \& Editing,
    Visualization.
\textbf{Jörg Main:}
    Conceptualization,
    Methodology,
    Formal analysis,
    Resources,
    Writing -- Original Draft,
    Writing -- Review \& Editing,
    Supervision,
    Project administration,
    Funding acquisition.
\textbf{Rigoberto Hernandez:}
    Conceptualization,
    Writing -- Review \& Editing,
    Project administration,
    Funding acquisition.


\section*{Acknowledgments}

Fruitful discussions with Robin Bardakcioglu, Matthias Feldmaier,
and Andrej Junginger are gratefully acknowledged.
The German portion of this collaborative work was supported
by Deutsche Forschungsgemeinschaft (DFG) through Grant No.\ MA1639/14-1.
RH's contribution to this work was supported
by the National Science Foundation (NSF) through Grant No.\ CHE-1700749.
This collaboration has also benefited from support
by the European Union's Horizon 2020 Research and Innovation Program
under the Marie Skłodowska-Curie Grant Agreement No.\ 734557.


\bibliographystyle{elsarticle-num-names}
\bibliography{paper-q15}

\begin{thebibliography}{71}
\expandafter\ifx\csname natexlab\endcsname\relax\def\natexlab#1{#1}\fi
\providecommand{\url}[1]{\texttt{#1}}
\providecommand{\href}[2]{#2}
\providecommand{\path}[1]{#1}
\providecommand{\DOIprefix}{doi:}
\providecommand{\ArXivprefix}{arXiv:}
\providecommand{\URLprefix}{URL: }
\providecommand{\Pubmedprefix}{pmid:}
\providecommand{\doi}[1]{\href{http://dx.doi.org/#1}{\path{#1}}}
\providecommand{\Pubmed}[1]{\href{pmid:#1}{\path{#1}}}
\providecommand{\bibinfo}[2]{#2}
\ifx\xfnm\relax \def\xfnm[#1]{\unskip,\space#1}\fi
\bibitem[{Schneider et~al.(2004)Schneider, Zhao, Kozhuharova, Groudeva-Zotova,
  M{\"u}hl, Ritschel, M{\"o}nch, Vinzelberg, Elefant, Graff
  et~al.}]{Schneider2004}
\bibinfo{author}{C.~M. Schneider}, \bibinfo{author}{B.~Zhao},
  \bibinfo{author}{R.~Kozhuharova}, \bibinfo{author}{S.~Groudeva-Zotova},
  \bibinfo{author}{T.~M{\"u}hl}, \bibinfo{author}{M.~Ritschel},
  \bibinfo{author}{I.~M{\"o}nch}, \bibinfo{author}{H.~Vinzelberg},
  \bibinfo{author}{D.~Elefant}, \bibinfo{author}{A.~Graff}, et~al.,
\newblock \bibinfo{title}{Towards molecular spintronics: magnetotransport and
  magnetism in carbon nanotube-based systems},
\newblock \bibinfo{journal}{Diam. Relat. Mater.} \bibinfo{volume}{13}
  (\bibinfo{year}{2004}) \bibinfo{pages}{215--220}.
  \DOIprefix\doi{10.1016/j.diamond.2003.10.009}.
\bibitem[{Jiang et~al.(2005)Jiang, Wang, Shelby, Macfarlane, Bank, Harris, and
  Parkin}]{Jiang2005}
\bibinfo{author}{X.~Jiang}, \bibinfo{author}{R.~Wang}, \bibinfo{author}{R.~M.
  Shelby}, \bibinfo{author}{R.~M. Macfarlane}, \bibinfo{author}{S.~R. Bank},
  \bibinfo{author}{J.~S. Harris}, \bibinfo{author}{S.~S.~P. Parkin},
\newblock \bibinfo{title}{Highly spin-polarized room-temperature tunnel
  injector for semiconductor spintronics using {MgO} (100)},
\newblock \bibinfo{journal}{Phys. Rev. Lett.} \bibinfo{volume}{94}
  (\bibinfo{year}{2005}) \bibinfo{pages}{056601}.
  \DOIprefix\doi{10.1103/PhysRevLett.94.056601}.
\bibitem[{Cowburn(2007)}]{Cowburn2007}
\bibinfo{author}{R.~P. Cowburn},
\newblock \bibinfo{title}{Spintronics: Change of direction},
\newblock \bibinfo{journal}{Nat. Mater.} \bibinfo{volume}{6}
  (\bibinfo{year}{2007}) \bibinfo{pages}{255--256}.
  \DOIprefix\doi{10.1038/nmat1877}.
\bibitem[{Shinjo(2014)}]{Shinjo2014}
\bibinfo{author}{T.~Shinjo},
\newblock \bibinfo{title}{Overview},
\newblock in: \bibinfo{booktitle}{Nanomagnetism and Spintronics},
  \bibinfo{edition}{second edition} ed., \bibinfo{publisher}{Elsevier},
  \bibinfo{year}{2014}, pp. \bibinfo{pages}{1--14}.
\bibitem[{Maekawa et~al.(2017)Maekawa, Valenzuela, Saitoh, and
  Kimura}]{Maekawa2017}
\bibinfo{editor}{S.~Maekawa}, \bibinfo{editor}{S.~O. Valenzuela},
  \bibinfo{editor}{E.~Saitoh}, \bibinfo{editor}{T.~Kimura} (Eds.),
  \bibinfo{title}{Spin current}, volume~\bibinfo{volume}{22},
  \bibinfo{publisher}{Oxford University Press}, \bibinfo{year}{2017}.
\bibitem[{Monakhov et~al.(2017)Monakhov, Moors, and
  K{\"o}gerler}]{Monakhov2017}
\bibinfo{author}{K.~Y. Monakhov}, \bibinfo{author}{M.~Moors},
  \bibinfo{author}{P.~K{\"o}gerler},
\newblock \bibinfo{title}{Chapter nine - perspectives for polyoxometalates in
  single-molecule electronics and spintronics},
\newblock in: \bibinfo{editor}{R.~van Eldik}, \bibinfo{editor}{L.~Cronin}
  (Eds.), \bibinfo{booktitle}{Polyoxometalate Chemistry},
  volume~\bibinfo{volume}{69} of \textit{\bibinfo{series}{Advances in Inorganic
  Chemistry}}, \bibinfo{publisher}{Academic Press}, \bibinfo{year}{2017}, pp.
  \bibinfo{pages}{251--286}. \DOIprefix\doi{10.1016/bs.adioch.2016.12.009}.
\bibitem[{Khodadadi et~al.(2020)Khodadadi, Rai, Sapkota, Srivastava, Nepal,
  Lim, Smith, Mewes, Budhathoki, Hauser et~al.}]{Khodadadi2020a}
\bibinfo{author}{B.~Khodadadi}, \bibinfo{author}{A.~Rai},
  \bibinfo{author}{A.~Sapkota}, \bibinfo{author}{A.~Srivastava},
  \bibinfo{author}{B.~Nepal}, \bibinfo{author}{Y.~Lim}, \bibinfo{author}{D.~A.
  Smith}, \bibinfo{author}{C.~Mewes}, \bibinfo{author}{S.~Budhathoki},
  \bibinfo{author}{A.~J. Hauser}, et~al.,
\newblock \bibinfo{title}{Conductivitylike {Gilbert} damping due to intraband
  scattering in epitaxial iron},
\newblock \bibinfo{journal}{Phys. Rev. Lett.} \bibinfo{volume}{124}
  (\bibinfo{year}{2020}) \bibinfo{pages}{157201}.
  \DOIprefix\doi{10.1103/physrevlett.124.157201}.
\bibitem[{Liu et~al.(2020)Liu, Yu, Gonz{\'{a}}lez-Hern{\'{a}}ndez, Li, Deng,
  Lin, Zhou, Zhou, Zhou, Wang, Guo, Yoong, Chow, Han, Dup{\'{e}},
  {\v{Z}}elezn{\'{y}}, Sinova, and Chen}]{Liu2020a}
\bibinfo{author}{L.~Liu}, \bibinfo{author}{J.~Yu},
  \bibinfo{author}{R.~Gonz{\'{a}}lez-Hern{\'{a}}ndez}, \bibinfo{author}{C.~Li},
  \bibinfo{author}{J.~Deng}, \bibinfo{author}{W.~Lin},
  \bibinfo{author}{C.~Zhou}, \bibinfo{author}{T.~Zhou},
  \bibinfo{author}{J.~Zhou}, \bibinfo{author}{H.~Wang},
  \bibinfo{author}{R.~Guo}, \bibinfo{author}{H.~Y. Yoong},
  \bibinfo{author}{G.~M. Chow}, \bibinfo{author}{X.~Han},
  \bibinfo{author}{B.~Dup{\'{e}}}, \bibinfo{author}{J.~{\v{Z}}elezn{\'{y}}},
  \bibinfo{author}{J.~Sinova}, \bibinfo{author}{J.~Chen},
\newblock \bibinfo{title}{Electrical switching of perpendicular magnetization
  in a single ferromagnetic layer},
\newblock \bibinfo{journal}{Phys. Rev. B} \bibinfo{volume}{101}
  (\bibinfo{year}{2020}) \bibinfo{pages}{220402}.
  \DOIprefix\doi{10.1103/physrevb.101.220402}.
\bibitem[{Wolf et~al.(2001)Wolf, Awschalom, Buhrman, Daughton, von Moln{\'a}r,
  Roukes, Chtchelkanova, and Treger}]{Wolf2001}
\bibinfo{author}{S.~A. Wolf}, \bibinfo{author}{D.~D. Awschalom},
  \bibinfo{author}{R.~A. Buhrman}, \bibinfo{author}{J.~M. Daughton},
  \bibinfo{author}{S.~von Moln{\'a}r}, \bibinfo{author}{M.~L. Roukes},
  \bibinfo{author}{A.~Y. Chtchelkanova}, \bibinfo{author}{D.~M. Treger},
\newblock \bibinfo{title}{Spintronics: A spin-based electronics vision for the
  future},
\newblock \bibinfo{journal}{Science} \bibinfo{volume}{294}
  (\bibinfo{year}{2001}) \bibinfo{pages}{1488--1495}.
  \DOIprefix\doi{10.1126/science.1065389}.
\bibitem[{Rocha et~al.(2005)Rocha, Garcia-Suarez, Bailey, Lambert, Ferrer, and
  Sanvito}]{Rocha2005}
\bibinfo{author}{A.~R. Rocha}, \bibinfo{author}{V.~M. Garcia-Suarez},
  \bibinfo{author}{S.~W. Bailey}, \bibinfo{author}{C.~J. Lambert},
  \bibinfo{author}{J.~Ferrer}, \bibinfo{author}{S.~Sanvito},
\newblock \bibinfo{title}{Towards molecular spintronics},
\newblock \bibinfo{journal}{Nat. Mater.} \bibinfo{volume}{4}
  (\bibinfo{year}{2005}) \bibinfo{pages}{335--339}.
  \DOIprefix\doi{10.1038/nmat1349}.
\bibitem[{Adam et~al.(2006)Adam, Polianski, and Brouwer}]{Adam2006}
\bibinfo{author}{S.~Adam}, \bibinfo{author}{M.~L. Polianski},
  \bibinfo{author}{P.~W. Brouwer},
\newblock \bibinfo{title}{Current-induced transverse spin-wave instability in
  thin ferromagnets: Beyond linear stability analysis},
\newblock \bibinfo{journal}{Phys. Rev. B} \bibinfo{volume}{73}
  (\bibinfo{year}{2006}) \bibinfo{pages}{024425}.
  \DOIprefix\doi{10.1103/PhysRevB.73.024425}.
\bibitem[{Chappert et~al.(2007)Chappert, Fert, and van Dau}]{Chappert2007}
\bibinfo{author}{C.~Chappert}, \bibinfo{author}{A.~Fert},
  \bibinfo{author}{F.~N. van Dau},
\newblock \bibinfo{title}{The emergence of spin electronics in data storage},
\newblock \bibinfo{journal}{Nat. Mater.} \bibinfo{volume}{6}
  (\bibinfo{year}{2007}) \bibinfo{pages}{813--823}.
  \DOIprefix\doi{10.1142/9789814287005\_0015}.
\bibitem[{Taniguchi et~al.(2013)Taniguchi, Utsumi, Marthaler, Golubev, and
  Imamura}]{Taniguchi2013b}
\bibinfo{author}{T.~Taniguchi}, \bibinfo{author}{Y.~Utsumi},
  \bibinfo{author}{M.~Marthaler}, \bibinfo{author}{D.~S. Golubev},
  \bibinfo{author}{H.~Imamura},
\newblock \bibinfo{title}{Spin torque switching of an in-plane magnetized
  system in a thermally activated region},
\newblock \bibinfo{journal}{Phys. Rev. B} \bibinfo{volume}{87}
  (\bibinfo{year}{2013}) \bibinfo{pages}{054406}.
  \DOIprefix\doi{10.1103/PhysRevB.87.054406}.
\bibitem[{Gilbert(2004)}]{Gilbert2004}
\bibinfo{author}{T.~L. Gilbert},
\newblock \bibinfo{title}{A phenomenological theory of damping in ferromagnetic
  materials},
\newblock \bibinfo{journal}{IEEE Trans. Magn.} \bibinfo{volume}{40}
  (\bibinfo{year}{2004}) \bibinfo{pages}{3443--3449}.
  \DOIprefix\doi{10.1109/TMAG.2004.836740}.
\bibitem[{Apalkov and Visscher(2005)}]{Apalkov2005}
\bibinfo{author}{D.~M. Apalkov}, \bibinfo{author}{P.~B. Visscher},
\newblock \bibinfo{title}{Spin-torque switching: {Fokker-Planck} rate
  calculation},
\newblock \bibinfo{journal}{Phys. Rev. B} \bibinfo{volume}{72}
  (\bibinfo{year}{2005}) \bibinfo{pages}{180405}.
  \DOIprefix\doi{10.1103/PhysRevB.72.180405}.
\bibitem[{Abert(2019)}]{Abert2019a}
\bibinfo{author}{C.~Abert},
\newblock \bibinfo{title}{Micromagnetics and spintronics: models and numerical
  methods},
\newblock \bibinfo{journal}{Eur. Phys. J. B} \bibinfo{volume}{92}
  (\bibinfo{year}{2019}) \bibinfo{pages}{120}.
  \DOIprefix\doi{10.1140/epjb/e2019-90599-6}.
\bibitem[{Li and Zhang(2003)}]{LiZhang2003}
\bibinfo{author}{Z.~Li}, \bibinfo{author}{S.~Zhang},
\newblock \bibinfo{title}{Magnetization dynamics with a spin-transfer torque},
\newblock \bibinfo{journal}{Phys. Rev. B} \bibinfo{volume}{68}
  (\bibinfo{year}{2003}) \bibinfo{pages}{024404}.
  \DOIprefix\doi{10.1103/PhysRevB.68.024404}.
\bibitem[{Li and Zhang(2004)}]{LiZhang2004}
\bibinfo{author}{Z.~Li}, \bibinfo{author}{S.~Zhang},
\newblock \bibinfo{title}{Thermally assisted magnetization reversal in the
  presence of a spin-transfer torque},
\newblock \bibinfo{journal}{Phys. Rev. B} \bibinfo{volume}{69}
  (\bibinfo{year}{2004}) \bibinfo{pages}{134416}.
  \DOIprefix\doi{10.1103/PhysRevB.69.134416}.
\bibitem[{Slonczewski(1996)}]{Slonczewski1996a}
\bibinfo{author}{J.~C. Slonczewski},
\newblock \bibinfo{title}{Current-driven excitation of magnetic multilayers},
\newblock \bibinfo{journal}{J. Magn. Magn. Mater.} \bibinfo{volume}{159}
  (\bibinfo{year}{1996}) \bibinfo{pages}{L1--L7}.
  \DOIprefix\doi{10.1016/0304-8853(96)00062-5}.
\bibitem[{Zhu et~al.(2008)Zhu, Zhu, and Tang}]{Zhu2008a}
\bibinfo{author}{J.-G. Zhu}, \bibinfo{author}{X.~Zhu},
  \bibinfo{author}{Y.~Tang},
\newblock \bibinfo{title}{Microwave assisted magnetic recording},
\newblock \bibinfo{journal}{IEEE Trans. Magn.} \bibinfo{volume}{44}
  (\bibinfo{year}{2008}) \bibinfo{pages}{125--131}.
  \DOIprefix\doi{10.1109/tmag.2007.911031}.
\bibitem[{Okamoto et~al.(2012)Okamoto, Kikuchi, Furuta, Kitakami, and
  Shimatsu}]{Okamoto2012a}
\bibinfo{author}{S.~Okamoto}, \bibinfo{author}{N.~Kikuchi},
  \bibinfo{author}{M.~Furuta}, \bibinfo{author}{O.~Kitakami},
  \bibinfo{author}{T.~Shimatsu},
\newblock \bibinfo{title}{Switching behaviors and its dynamics of a {Co}/{Pt}
  nanodot under the assistance of rf fields},
\newblock \bibinfo{journal}{Phys. Rev. Lett.} \bibinfo{volume}{109}
  (\bibinfo{year}{2012}). \DOIprefix\doi{10.1103/physrevlett.109.237209}.
\bibitem[{Taniguchi(2014)}]{Taniguchi2014b}
\bibinfo{author}{T.~Taniguchi},
\newblock \bibinfo{title}{Magnetization reversal condition for a nanomagnet
  within a rotating magnetic field},
\newblock \bibinfo{journal}{Phys. Rev. B} \bibinfo{volume}{90}
  (\bibinfo{year}{2014}). \DOIprefix\doi{10.1103/physrevb.90.024424}.
\bibitem[{Suto et~al.(2015)Suto, Nagasawa, Kudo, Mizushima, and
  Sato}]{Suto2015a}
\bibinfo{author}{H.~Suto}, \bibinfo{author}{T.~Nagasawa},
  \bibinfo{author}{K.~Kudo}, \bibinfo{author}{K.~Mizushima},
  \bibinfo{author}{R.~Sato},
\newblock \bibinfo{title}{Microwave-assisted switching of a single
  perpendicular magnetic tunnel junction nanodot},
\newblock \bibinfo{journal}{Appl. Phys. Express} \bibinfo{volume}{8}
  (\bibinfo{year}{2015}) \bibinfo{pages}{023001}.
  \DOIprefix\doi{10.7567/apex.8.023001}.
\bibitem[{Barros et~al.(2011)Barros, Rassam, Jirari, and
  Kachkachi}]{Barros2011a}
\bibinfo{author}{N.~Barros}, \bibinfo{author}{M.~Rassam},
  \bibinfo{author}{H.~Jirari}, \bibinfo{author}{H.~Kachkachi},
\newblock \bibinfo{title}{Optimal switching of a nanomagnet assisted by
  microwaves},
\newblock \bibinfo{journal}{Phys. Rev. B} \bibinfo{volume}{83}
  (\bibinfo{year}{2011}). \DOIprefix\doi{10.1103/physrevb.83.144418}.
\bibitem[{Barros et~al.(2013)Barros, Rassam, and Kachkachi}]{Barros2013a}
\bibinfo{author}{N.~Barros}, \bibinfo{author}{H.~Rassam},
  \bibinfo{author}{H.~Kachkachi},
\newblock \bibinfo{title}{Microwave-assisted switching of a nanomagnet:
  Analytical determination of the optimal microwave field},
\newblock \bibinfo{journal}{Phys. Rev. B} \bibinfo{volume}{88}
  (\bibinfo{year}{2013}). \DOIprefix\doi{10.1103/physrevb.88.014421}.
\bibitem[{Klughertz et~al.(2015)Klughertz, Friedland, Hervieux, and
  Manfredi}]{Klughertz2015a}
\bibinfo{author}{G.~Klughertz}, \bibinfo{author}{L.~Friedland},
  \bibinfo{author}{P.-A. Hervieux}, \bibinfo{author}{G.~Manfredi},
\newblock \bibinfo{title}{Autoresonant switching of the magnetization in
  single-domain nanoparticles: Two-level theory},
\newblock \bibinfo{journal}{Phys. Rev. B} \bibinfo{volume}{91}
  (\bibinfo{year}{2015}). \DOIprefix\doi{10.1103/physrevb.91.104433}.
\bibitem[{Taniguchi et~al.(2016)Taniguchi, Saida, Nakatani, and
  Kubota}]{Taniguchi2016a}
\bibinfo{author}{T.~Taniguchi}, \bibinfo{author}{D.~Saida},
  \bibinfo{author}{Y.~Nakatani}, \bibinfo{author}{H.~Kubota},
\newblock \bibinfo{title}{Magnetization switching by current and microwaves},
\newblock \bibinfo{journal}{Phys. Rev. B} \bibinfo{volume}{93}
  (\bibinfo{year}{2016}). \DOIprefix\doi{10.1103/physrevb.93.014430}.
\bibitem[{Rivkin and Ketterson(2006)}]{Rivkin2006a}
\bibinfo{author}{K.~Rivkin}, \bibinfo{author}{J.~B. Ketterson},
\newblock \bibinfo{title}{Magnetization reversal in the anisotropy-dominated
  regime using time-dependent magnetic fields},
\newblock \bibinfo{journal}{Appl. Phys. Lett.} \bibinfo{volume}{89}
  (\bibinfo{year}{2006}) \bibinfo{pages}{252507}.
  \DOIprefix\doi{10.1063/1.2405855}.
\bibitem[{Taniguchi(2015)}]{Taniguchi2015a}
\bibinfo{author}{T.~Taniguchi},
\newblock \bibinfo{title}{Magnetization switching by microwaves synchronized in
  the vicinity of precession frequency},
\newblock \bibinfo{journal}{Appl. Phys. Express} \bibinfo{volume}{8}
  (\bibinfo{year}{2015}) \bibinfo{pages}{083004}.
  \DOIprefix\doi{10.7567/apex.8.083004}.
\bibitem[{Eyring(1935)}]{eyring35}
\bibinfo{author}{H.~Eyring},
\newblock \bibinfo{title}{The activated complex in chemical reactions},
\newblock \bibinfo{journal}{J. Chem. Phys.} \bibinfo{volume}{3}
  (\bibinfo{year}{1935}) \bibinfo{pages}{107--115}.
  \DOIprefix\doi{10.1063/1.1749604}.
\bibitem[{Wigner(1937)}]{wigner37}
\bibinfo{author}{E.~P. Wigner},
\newblock \bibinfo{title}{Calculation of the rate of elementary association
  reactions},
\newblock \bibinfo{journal}{J. Chem. Phys.} \bibinfo{volume}{5}
  (\bibinfo{year}{1937}) \bibinfo{pages}{720--725}.
  \DOIprefix\doi{10.1063/1.1750107}.
\bibitem[{Pitzer et~al.(1962)Pitzer, Smith, and Eyring}]{pitzer}
\bibinfo{author}{K.~S. Pitzer}, \bibinfo{author}{F.~T. Smith},
  \bibinfo{author}{H.~Eyring}, \bibinfo{title}{The Transition State}, Special
  Publ., \bibinfo{publisher}{Chemical Society}, \bibinfo{address}{London},
  \bibinfo{year}{1962}.
\bibitem[{Pechukas(1981)}]{pech81}
\bibinfo{author}{P.~Pechukas},
\newblock \bibinfo{title}{Transition state theory},
\newblock \bibinfo{journal}{Annu. Rev. Phys. Chem.} \bibinfo{volume}{32}
  (\bibinfo{year}{1981}) \bibinfo{pages}{159--177}.
  \DOIprefix\doi{10.1146/annurev.pc.32.100181.001111}.
\bibitem[{Truhlar et~al.(1996)Truhlar, Garrett, and Klippenstein}]{truh96}
\bibinfo{author}{D.~G. Truhlar}, \bibinfo{author}{B.~C. Garrett},
  \bibinfo{author}{S.~J. Klippenstein},
\newblock \bibinfo{title}{Current status of transition-state theory},
\newblock \bibinfo{journal}{J. Phys. Chem.} \bibinfo{volume}{100}
  (\bibinfo{year}{1996}) \bibinfo{pages}{12771--12800}.
  \DOIprefix\doi{10.1021/jp953748q}.
\bibitem[{Mullen et~al.(2014)Mullen, Shea, and Peters}]{peters14a}
\bibinfo{author}{R.~G. Mullen}, \bibinfo{author}{J.-E. Shea},
  \bibinfo{author}{B.~Peters},
\newblock \bibinfo{title}{Communication: An existence test for dividing
  surfaces without recrossing},
\newblock \bibinfo{journal}{J. Chem. Phys.} \bibinfo{volume}{140}
  (\bibinfo{year}{2014}) \bibinfo{pages}{041104}.
  \DOIprefix\doi{10.1063/1.4862504}.
\bibitem[{Wiggins(2016)}]{wiggins16}
\bibinfo{author}{S.~Wiggins},
\newblock \bibinfo{title}{The role of normally hyperbolic invariant manifolds
  ({NHIMS}) in the context of the phase space setting for chemical reaction
  dynamics},
\newblock \bibinfo{journal}{Regul. Chaotic Dyn.} \bibinfo{volume}{21}
  (\bibinfo{year}{2016}) \bibinfo{pages}{621--638}.
  \DOIprefix\doi{10.1134/S1560354716060034}.
\bibitem[{Jaff{\'e} et~al.(2000)Jaff{\'e}, Farrelly, and Uzer}]{Jaffe00}
\bibinfo{author}{C.~Jaff{\'e}}, \bibinfo{author}{D.~Farrelly},
  \bibinfo{author}{T.~Uzer},
\newblock \bibinfo{title}{Transition state theory without time-reversal
  symmetry: Chaotic ionization of the hydrogen atom},
\newblock \bibinfo{journal}{Phys. Rev. Lett.} \bibinfo{volume}{84}
  (\bibinfo{year}{2000}) \bibinfo{pages}{610--613}.
  \DOIprefix\doi{10.1103/PhysRevLett.84.610}.
\bibitem[{Jacucci et~al.(1984)Jacucci, Toller, DeLorenzi, and
  Flynn}]{Jacucci1984}
\bibinfo{author}{G.~Jacucci}, \bibinfo{author}{M.~Toller},
  \bibinfo{author}{G.~DeLorenzi}, \bibinfo{author}{C.~P. Flynn},
\newblock \bibinfo{title}{{Rate Theory, Return Jump Catastrophes, and the
  Center Manifold}},
\newblock \bibinfo{journal}{Phys. Rev. Lett.} \bibinfo{volume}{52}
  (\bibinfo{year}{1984}) \bibinfo{pages}{295}.
  \DOIprefix\doi{10.1103/PhysRevLett.52.295}.
\bibitem[{Komatsuzaki and Berry(1999)}]{KomatsuzakiBerry99a}
\bibinfo{author}{T.~Komatsuzaki}, \bibinfo{author}{R.~S. Berry},
\newblock \bibinfo{title}{Regularity in chaotic reaction paths. {I.} {${\rm
  Ar}_6$}},
\newblock \bibinfo{journal}{J. Chem. Phys.} \bibinfo{volume}{110}
  (\bibinfo{year}{1999}) \bibinfo{pages}{9160--9173}.
  \DOIprefix\doi{10.1063/1.478838}.
\bibitem[{Komatsuzaki and Berry(2002)}]{KomatsuzakiBerry02}
\bibinfo{author}{T.~Komatsuzaki}, \bibinfo{author}{R.~S. Berry},
\newblock \bibinfo{title}{Chemical reaction dynamics: Many-body chaos and
  regularity},
\newblock \bibinfo{journal}{Adv. Chem. Phys.} \bibinfo{volume}{123}
  (\bibinfo{year}{2002}) \bibinfo{pages}{79--152}.
  \DOIprefix\doi{10.1002/0471231509.ch2}.
\bibitem[{Toller et~al.(1985)Toller, Jacucci, DeLorenzi, and Flynn}]{toller}
\bibinfo{author}{M.~Toller}, \bibinfo{author}{G.~Jacucci},
  \bibinfo{author}{G.~DeLorenzi}, \bibinfo{author}{C.~P. Flynn},
\newblock \bibinfo{title}{Theory of classical diffusion jumps in solids},
\newblock \bibinfo{journal}{Phys. Rev. B} \bibinfo{volume}{32}
  (\bibinfo{year}{1985}) \bibinfo{pages}{2082}.
  \DOIprefix\doi{10.1103/PhysRevB.32.2082}.
\bibitem[{Voter et~al.(2002)Voter, Montalenti, and Germann}]{voter02b}
\bibinfo{author}{A.~F. Voter}, \bibinfo{author}{F.~Montalenti},
  \bibinfo{author}{T.~C. Germann},
\newblock \bibinfo{title}{Extending the time scale in atomistic simulations of
  materials},
\newblock \bibinfo{journal}{Annu.~Rev.~Mater.~Res.} \bibinfo{volume}{32}
  (\bibinfo{year}{2002}) \bibinfo{pages}{321--346}.
  \DOIprefix\doi{10.1146/annurev.matsci.32.112601.141541}.
\bibitem[{de~Oliveira et~al.(2002)de~Oliveira, {Ozorio de Almeida}, {Dami{\~a}o
  Soares}, and Tonini}]{Oliveira02}
\bibinfo{author}{H.~P. de~Oliveira}, \bibinfo{author}{A.~M. {Ozorio de
  Almeida}}, \bibinfo{author}{I.~{Dami{\~a}o Soares}}, \bibinfo{author}{E.~V.
  Tonini},
\newblock \bibinfo{title}{Homoclinic chaos in the dynamics of a general
  {Bianchi} type-{IX} model},
\newblock \bibinfo{journal}{Phys. Rev. D} \bibinfo{volume}{65}
  (\bibinfo{year}{2002}) \bibinfo{pages}{083511/1--9}.
  \DOIprefix\doi{10.1103/PhysRevD.65.083511}.
\bibitem[{Jaff{\'e} et~al.(2002)Jaff{\'e}, Ross, Lo, Marsden, Farrelly, and
  Uzer}]{Jaffe02}
\bibinfo{author}{C.~Jaff{\'e}}, \bibinfo{author}{S.~D. Ross},
  \bibinfo{author}{M.~W. Lo}, \bibinfo{author}{J.~Marsden},
  \bibinfo{author}{D.~Farrelly}, \bibinfo{author}{T.~Uzer},
\newblock \bibinfo{title}{Statistical theory of asteroid escape rates},
\newblock \bibinfo{journal}{Phys. Rev. Lett.} \bibinfo{volume}{89}
  (\bibinfo{year}{2002}) \bibinfo{pages}{011101}.
  \DOIprefix\doi{10.1103/PhysRevLett.89.011101}.
\bibitem[{Stallings et~al.(2020)Stallings, Iyer, and Hernandez}]{hern20i}
\bibinfo{author}{D.~Stallings}, \bibinfo{author}{S.~K. Iyer},
  \bibinfo{author}{R.~Hernandez},
\newblock \bibinfo{title}{Removing barriers},
\newblock in: \bibinfo{editor}{S.~Azad} (Ed.), \bibinfo{booktitle}{Addressing
  Gender Bias in Science \& Technology}, volume \bibinfo{volume}{1354} of
  \textit{\bibinfo{series}{ACS Symposium Series}}, \bibinfo{publisher}{American
  Chemical Society; Oxford University Press}, \bibinfo{address}{Washington DC},
  \bibinfo{year}{2020}, pp. \bibinfo{pages}{91--108}.
  \DOIprefix\doi{10.1021/bk-2020-1354.ch006}.
\bibitem[{Bartsch et~al.(2008)Bartsch, Moix, Hernandez, Kawai, and
  Uzer}]{hern08d}
\bibinfo{author}{T.~Bartsch}, \bibinfo{author}{J.~M. Moix},
  \bibinfo{author}{R.~Hernandez}, \bibinfo{author}{S.~Kawai},
  \bibinfo{author}{T.~Uzer},
\newblock \bibinfo{title}{Time-dependent transition state theory},
\newblock \bibinfo{journal}{Adv. Chem. Phys.} \bibinfo{volume}{140}
  (\bibinfo{year}{2008}) \bibinfo{pages}{191--238}.
  \DOIprefix\doi{10.1002/9780470371572.ch4}.
\bibitem[{Hernandez and Miller(1993)}]{hern93b}
\bibinfo{author}{R.~Hernandez}, \bibinfo{author}{W.~H. Miller},
\newblock \bibinfo{title}{Semiclassical transition state theory. {A} new
  perspective},
\newblock \bibinfo{journal}{Chem. Phys. Lett.} \bibinfo{volume}{214}
  (\bibinfo{year}{1993}) \bibinfo{pages}{129--136}.
  \DOIprefix\doi{10.1016/0009-2614(93)90071-8}.
\bibitem[{Uzer et~al.(2002)Uzer, Jaff\'e, Palaci{\'a}n, Yanguas, and
  Wiggins}]{Uzer02}
\bibinfo{author}{T.~Uzer}, \bibinfo{author}{C.~Jaff\'e},
  \bibinfo{author}{J.~Palaci{\'a}n}, \bibinfo{author}{P.~Yanguas},
  \bibinfo{author}{S.~Wiggins},
\newblock \bibinfo{title}{The geometry of reaction dynamics},
\newblock \bibinfo{journal}{Nonlinearity} \bibinfo{volume}{15}
  (\bibinfo{year}{2002}) \bibinfo{pages}{957--992}.
  \DOIprefix\doi{10.1088/0951-7715/15/4/301}.
\bibitem[{Waalkens and Wiggins(2004)}]{Waalkens04b}
\bibinfo{author}{H.~Waalkens}, \bibinfo{author}{S.~Wiggins},
\newblock \bibinfo{title}{Direct construction of a dividing surface of minimal
  flux for multi-degree-of-freedom systems that cannot be recrossed},
\newblock \bibinfo{journal}{J. Phys. A} \bibinfo{volume}{37}
  (\bibinfo{year}{2004}) \bibinfo{pages}{L435--L445}.
  \DOIprefix\doi{10.1088/0305-4470/37/35/L02}.
\bibitem[{\ifmmode \mbox{\c{C}}\else \c{C}\fi{}ift\ifmmode~\mbox{\c{c}}\else
  \c{c}\fi{}i and Waalkens(2013)}]{Waalkens13}
\bibinfo{author}{U.~\ifmmode \mbox{\c{C}}\else
  \c{C}\fi{}ift\ifmmode~\mbox{\c{c}}\else \c{c}\fi{}i},
  \bibinfo{author}{H.~Waalkens},
\newblock \bibinfo{title}{Reaction dynamics through kinetic transition states},
\newblock \bibinfo{journal}{Phys. Rev. Lett.} \bibinfo{volume}{110}
  (\bibinfo{year}{2013}) \bibinfo{pages}{233201}.
  \DOIprefix\doi{10.1103/PhysRevLett.110.233201}.
\bibitem[{Craven et~al.(2014)Craven, Bartsch, and Hernandez}]{hern14f}
\bibinfo{author}{G.~T. Craven}, \bibinfo{author}{T.~Bartsch},
  \bibinfo{author}{R.~Hernandez},
\newblock \bibinfo{title}{Communication: Transition state trajectory stability
  determines barrier crossing rates in chemical reactions induced by
  time-dependent oscillating fields},
\newblock \bibinfo{journal}{J. Chem. Phys.} \bibinfo{volume}{141}
  (\bibinfo{year}{2014}) \bibinfo{pages}{041106}.
  \DOIprefix\doi{10.1063/1.4891471}.
\bibitem[{Feldmaier et~al.(2017)Feldmaier, Junginger, Main, Wunner, and
  Hernandez}]{hern17h}
\bibinfo{author}{M.~Feldmaier}, \bibinfo{author}{A.~Junginger},
  \bibinfo{author}{J.~Main}, \bibinfo{author}{G.~Wunner},
  \bibinfo{author}{R.~Hernandez},
\newblock \bibinfo{title}{Obtaining time-dependent multi-dimensional dividing
  surfaces using {L}agrangian descriptors},
\newblock \bibinfo{journal}{Chem. Phys. Lett.} \bibinfo{volume}{687}
  (\bibinfo{year}{2017}) \bibinfo{pages}{194}.
  \DOIprefix\doi{10.1016/j.cplett.2017.09.008}.
\bibitem[{Feldmaier et~al.(2019{\natexlab{a}})Feldmaier, Schraft, Bardakcioglu,
  Reiff, Lober, Tsch{\"o}pe, Junginger, Main, Bartsch, and Hernandez}]{hern19a}
\bibinfo{author}{M.~Feldmaier}, \bibinfo{author}{P.~Schraft},
  \bibinfo{author}{R.~Bardakcioglu}, \bibinfo{author}{J.~Reiff},
  \bibinfo{author}{M.~Lober}, \bibinfo{author}{M.~Tsch{\"o}pe},
  \bibinfo{author}{A.~Junginger}, \bibinfo{author}{J.~Main},
  \bibinfo{author}{T.~Bartsch}, \bibinfo{author}{R.~Hernandez},
\newblock \bibinfo{title}{Invariant manifolds and rate constants in driven
  chemical reactions},
\newblock \bibinfo{journal}{J. Phys. Chem. B} \bibinfo{volume}{123}
  (\bibinfo{year}{2019}{\natexlab{a}}) \bibinfo{pages}{2070--2086}.
  \DOIprefix\doi{10.1021/acs.jpcb.8b10541}.
\bibitem[{Feldmaier et~al.(2019{\natexlab{b}})Feldmaier, Bardakcioglu, Reiff,
  Main, and Hernandez}]{hern19e}
\bibinfo{author}{M.~Feldmaier}, \bibinfo{author}{R.~Bardakcioglu},
  \bibinfo{author}{J.~Reiff}, \bibinfo{author}{J.~Main},
  \bibinfo{author}{R.~Hernandez},
\newblock \bibinfo{title}{Phase-space resolved rates in driven multidimensional
  chemical reactions},
\newblock \bibinfo{journal}{J. Chem. Phys.} \bibinfo{volume}{151}
  (\bibinfo{year}{2019}{\natexlab{b}}) \bibinfo{pages}{244108}.
  \DOIprefix\doi{10.1063/1.5127539}.
\bibitem[{Feldmaier et~al.(2020)Feldmaier, Reiff, Benito, Borondo, Main, and
  Hernandez}]{hern20m}
\bibinfo{author}{M.~Feldmaier}, \bibinfo{author}{J.~Reiff},
  \bibinfo{author}{R.~M. Benito}, \bibinfo{author}{F.~Borondo},
  \bibinfo{author}{J.~Main}, \bibinfo{author}{R.~Hernandez},
\newblock \bibinfo{title}{Influence of external driving on decays in the
  geometry of the {LiCN} isomerization},
\newblock \bibinfo{journal}{J. Chem. Phys.} \bibinfo{volume}{153}
  (\bibinfo{year}{2020}) \bibinfo{pages}{084115}.
  \DOIprefix\doi{10.1063/5.0015509}.
\bibitem[{Landau and Lifshitz(1935)}]{Landau1935}
\bibinfo{author}{L.~D. Landau}, \bibinfo{author}{E.~Lifshitz},
\newblock \bibinfo{title}{On the theory of the dispersion of magnetic
  permeability in ferromagnetic bodies},
\newblock \bibinfo{journal}{Phys. Z. Sowjetunion} \bibinfo{volume}{8}
  (\bibinfo{year}{1935}) \bibinfo{pages}{101--114}.
\bibitem[{Gilbert(1956)}]{Gilbert1956}
\bibinfo{author}{T.~L. Gilbert}, \bibinfo{title}{Formulation, Foundations and
  Applications of the Phenomenological Theory of Ferromagnetism.}, Ph.D.
  thesis, Illinois Institute of Technology, \bibinfo{year}{1956}.
\bibitem[{Ciornei(2010)}]{Ciornei2010}
\bibinfo{author}{M.-C. Ciornei}, \bibinfo{title}{Role of magnetic inertia in
  damped macrospin dynamics}, Ph.D. thesis, Ecole Polytechnique X,
  \bibinfo{year}{2010}. \URLprefix
  \url{https://pastel.archives-ouvertes.fr/tel-00460905}.
\bibitem[{Stoner and Wohlfarth(1948)}]{Stoner1948a}
\bibinfo{author}{E.~C. Stoner}, \bibinfo{author}{E.~P. Wohlfarth},
\newblock \bibinfo{title}{A mechanism of magnetic hysteresis in heterogeneous
  alloys},
\newblock \bibinfo{journal}{Philos. Trans. R. Soc. A} \bibinfo{volume}{240}
  (\bibinfo{year}{1948}) \bibinfo{pages}{599--642}.
  \DOIprefix\doi{10.1098/rsta.1948.0007}.
\bibitem[{Tannous and Gieraltowski(2008)}]{Tannous2008a}
\bibinfo{author}{C.~Tannous}, \bibinfo{author}{J.~Gieraltowski},
\newblock \bibinfo{title}{The {Stoner}--{Wohlfarth} model of ferromagnetism},
\newblock \bibinfo{journal}{Eur. J. Phys.} \bibinfo{volume}{29}
  (\bibinfo{year}{2008}) \bibinfo{pages}{475--487}.
  \DOIprefix\doi{10.1088/0143-0807/29/3/008}.
\bibitem[{Zhang and Li(2004)}]{Zhang2004a}
\bibinfo{author}{S.~Zhang}, \bibinfo{author}{Z.~Li},
\newblock \bibinfo{title}{Roles of nonequilibrium conduction electrons on the
  magnetization dynamics of ferromagnets},
\newblock \bibinfo{journal}{Phys. Rev. Lett.} \bibinfo{volume}{93}
  (\bibinfo{year}{2004}) \bibinfo{pages}{127204}.
  \DOIprefix\doi{10.1103/PhysRevLett.93.127204}.
\bibitem[{Manchon and Zhang(2008)}]{Manchon2008a}
\bibinfo{author}{A.~Manchon}, \bibinfo{author}{S.~Zhang},
\newblock \bibinfo{title}{Theory of nonequilibrium intrinsic spin torque in a
  single nanomagnet},
\newblock \bibinfo{journal}{Phys. Rev. B} \bibinfo{volume}{78}
  (\bibinfo{year}{2008}) \bibinfo{pages}{212405}.
  \DOIprefix\doi{10.1103/PhysRevB.78.212405}.
\bibitem[{McMichael et~al.(1999)McMichael, Donahue, Porter, and
  Eicke}]{McMichael1999a}
\bibinfo{author}{R.~D. McMichael}, \bibinfo{author}{M.~J. Donahue},
  \bibinfo{author}{D.~G. Porter}, \bibinfo{author}{J.~Eicke},
\newblock \bibinfo{title}{Comparison of magnetostatic field calculation methods
  on two-dimensional square grids as applied to a micromagnetic standard
  problem},
\newblock \bibinfo{journal}{J. Appl. Phys.} \bibinfo{volume}{85}
  (\bibinfo{year}{1999}) \bibinfo{pages}{5816--5818}.
  \DOIprefix\doi{10.1063/1.369929}.
\bibitem[{Zhu and Zhao(2020)}]{Zhu2020a}
\bibinfo{author}{D.~Zhu}, \bibinfo{author}{W.~Zhao},
\newblock \bibinfo{title}{Threshold current density for perpendicular
  magnetization switching through spin-orbit torque},
\newblock \bibinfo{journal}{Phys. Rev. Appl.} \bibinfo{volume}{13}
  (\bibinfo{year}{2020}) \bibinfo{pages}{044078}.
  \DOIprefix\doi{10.1103/PhysRevApplied.13.044078}.
\bibitem[{Apalkov and Visscher(2005)}]{Apalkov2005b}
\bibinfo{author}{D.~M. Apalkov}, \bibinfo{author}{P.~B. Visscher},
\newblock \bibinfo{title}{{Slonczewski} spin-torque as negative damping:
  {Fokker}--{Planck} computation of energy distribution},
\newblock \bibinfo{journal}{J. Magn. Magn. Mater.} \bibinfo{volume}{286}
  (\bibinfo{year}{2005}) \bibinfo{pages}{370--374}.
  \DOIprefix\doi{10.1016/j.jmmm.2004.09.094}.
\bibitem[{Najafi et~al.(2009)Najafi, Kr{\"u}ger, Bohlens, Franchin, Fangohr,
  Vanhaverbeke, Allenspach, Bolte, Merkt, Pfannkuche et~al.}]{Najafi2009a}
\bibinfo{author}{M.~Najafi}, \bibinfo{author}{B.~Kr{\"u}ger},
  \bibinfo{author}{S.~Bohlens}, \bibinfo{author}{M.~Franchin},
  \bibinfo{author}{H.~Fangohr}, \bibinfo{author}{A.~Vanhaverbeke},
  \bibinfo{author}{R.~Allenspach}, \bibinfo{author}{M.~Bolte},
  \bibinfo{author}{U.~Merkt}, \bibinfo{author}{D.~Pfannkuche}, et~al.,
\newblock \bibinfo{title}{Proposal for a standard problem for micromagnetic
  simulations including spin-transfer torque},
\newblock \bibinfo{journal}{J. Appl. Phys.} \bibinfo{volume}{105}
  (\bibinfo{year}{2009}) \bibinfo{pages}{113914}.
  \DOIprefix\doi{10.1063/1.3126702}.
\bibitem[{Glasstone et~al.(1941)Glasstone, Laidler, and Eyring}]{Glasstone1941}
\bibinfo{author}{S.~Glasstone}, \bibinfo{author}{K.~J. Laidler},
  \bibinfo{author}{H.~Eyring}, \bibinfo{title}{The Theory of Rate Processes:
  The Knetics of Chemical Reactions, Viscosity, Diffusion and Electrochemical
  Phenomena}, \bibinfo{publisher}{McGraw-Hill}, \bibinfo{address}{New York},
  \bibinfo{year}{1941}.
\bibitem[{Bardakcioglu et~al.(2018)Bardakcioglu, Junginger, Feldmaier, Main,
  and Hernandez}]{hern18g}
\bibinfo{author}{R.~Bardakcioglu}, \bibinfo{author}{A.~Junginger},
  \bibinfo{author}{M.~Feldmaier}, \bibinfo{author}{J.~Main},
  \bibinfo{author}{R.~Hernandez},
\newblock \bibinfo{title}{Binary contraction method for the construction of
  time-dependent dividing surfaces in driven chemical reactions},
\newblock \bibinfo{journal}{Phys. Rev. E} \bibinfo{volume}{98}
  (\bibinfo{year}{2018}) \bibinfo{pages}{032204}.
  \DOIprefix\doi{10.1103/PhysRevE.98.032204}.
\bibitem[{Junginger and Hernandez(2016)}]{hern16i}
\bibinfo{author}{A.~Junginger}, \bibinfo{author}{R.~Hernandez},
\newblock \bibinfo{title}{Lagrangian descriptors in dissipative systems},
\newblock \bibinfo{journal}{Phys. Chem. Chem. Phys.} \bibinfo{volume}{18}
  (\bibinfo{year}{2016}) \bibinfo{pages}{30282}.
  \DOIprefix\doi{10.1039/C6CP02532C}.
\bibitem[{Pollak et~al.(1989)Pollak, Grabert, and H{\"a}nggi}]{pgh89}
\bibinfo{author}{E.~Pollak}, \bibinfo{author}{H.~Grabert},
  \bibinfo{author}{P.~H{\"a}nggi},
\newblock \bibinfo{title}{Theory of activated rate processes for arbitrary
  frequency dependent friction: Solution of the turnover problem},
\newblock \bibinfo{journal}{J. Chem. Phys.} \bibinfo{volume}{91}
  (\bibinfo{year}{1989}) \bibinfo{pages}{4073--4087}.
  \DOIprefix\doi{10.1063/1.456837}.
\bibitem[{F{\"a}hnle et~al.(2011)F{\"a}hnle, Steiauf, and Illg}]{Fahnle2011}
\bibinfo{author}{M.~F{\"a}hnle}, \bibinfo{author}{D.~Steiauf},
  \bibinfo{author}{C.~Illg},
\newblock \bibinfo{title}{Generalized {G}ilbert equation including inertial
  damping: Derivation from an extended breathing {F}ermi surface model},
\newblock \bibinfo{journal}{Phys. Rev. B} \bibinfo{volume}{84}
  (\bibinfo{year}{2011}) \bibinfo{pages}{172403}.
  \DOIprefix\doi{10.1103/PhysRevB.84.172403}.

\end{thebibliography}
\end{document}